\documentclass{aa}  
\usepackage{graphicx}
\usepackage{txfonts}
\usepackage{multirow}
\usepackage{natbib}
\usepackage{mathrsfs}
\usepackage[normalem]{ulem}
\usepackage{adjustbox}
\usepackage{hyperref}
\usepackage[dvipsnames]{xcolor}

\begin{document} 

   \title{{Dancing on the Grain: Variety of CO and its isotopologue fluxes as a result of surface chemistry and T Tauri disk properties}}
   \titlerunning{CO and its isotopologue fluxes in T Tauri disks of varying parameters}

   \author{L. Zwicky
          \inst{1{,2}}\fnmsep\thanks{Corresponding author}
          \and
          T. Molyarova\inst{{3}}
          \and
          \'A. K\'osp\'al\inst{1,2,4}
          \and
          P. \'Abrah\'am\inst{1,2,5}
          }

   \institute{Konkoly Observatory, HUN-REN Research Centre for Astronomy and Earth Sciences, MTA Centre of Excellence, Konkoly-Thege Mikl\'os \'ut 15-17, 1121 Budapest, Hungary\\
              \email{lis.zwicky@csfk.org}
        \and
            ELTE E\"otv\"os Lor\'and University, Institute of Physics {and Astronomy}, P\'azm\'any P\'eter s\'et\'any 1/A, 1117 Budapest, Hungary
        \and
             School of Physics and Astronomy, University of Leeds, Leeds LS2 9JT, UK
        \and
            Max Planck Institute for Astronomy, K\"onigstuhl 17, 69117 Heidelberg, Germany
        \and
        Institute for Astronomy, University of Vienna, T\"urkenschanzstrasse 17, A-1180 Vienna, Austria
             }

   \date{Received December 12, 2024 / Accepted June 17, 2025}
 
  \abstract
   {One of the most important problems in the study of protoplanetary disks is the determination of their parameters, such as their size, age, stellar characteristics, and, most importantly, gas mass in the disk. At the moment, one of the main ways to infer the disk mass is to use a combination of CO isotopologue line observations. A number of theoretical studies have concluded that CO must be a reliable gas tracer as its relative abundance depends on disk parameters only weakly. However, the observed line fluxes cannot always be easily used to infer the column density, much less the abundance of CO.}
   {The aim of this work is to study the dependence of the CO isotopologue millimeter line fluxes on the astrochemical model parameters of a standard protoplanetary disk around a T Tauri star and to conclude whether they or their combinations can be reliably used to determine disk parameters. Our case is set apart from earlier studies in the literature by the usage of a comprehensive chemical network with grain surface chemistry together with line radiative transfer.}
   {We use the astrochemical model ANDES together with the radiative transfer code RADMC-3D to simulate CO isotopologue line fluxes from a set of disks with varying key parameters (disk mass, disk radius, stellar mass, and inclination). We study how these values change with one parameter varying and others fixed and approximate the dependences log-linearly.}
   {We describe the dependences of CO isotopologue fluxes on all chosen disk parameters. Physical and chemical processes responsible for these dependences are analyzed and explained for each parameter. We show {that using a combination of the $^{13}$CO and C$^{{18}}$O line fluxes,} the mass can be estimated only with{in} two orders of magnitude uncertainty {and characteristic radius within one order of magnitude uncertainty. We find that inclusion of grain surface chemistry reduces $^{13}$CO and C$^{18}$O fluxes which can explain the underestimation of disk mass in the previous studies.}}
   {}


   \maketitle
   
\defcitealias{2014ApJ...788...59W}{WB14}
\defcitealias{2016A&A...594A..85M}{M16}

\section{Introduction}

A crucial part of understanding protoplanetary disks and their evolution is our ability to describe them and their diversity. This is typically done through a set of key parameters that define the geometrical structure of the disk and the physical conditions throughout it. {Arguably, the most important parameter of the protoplanetary disk is its mass}. The disk mass influences the amount of matter potentially available for planet formation, as well as the contribution of gravitational instability to the dynamical evolution of the disk. In general, the mass of the disk is considered to be one of the key parameters in planet formation scenarios~\citep{2012A&A...541A..97M, 2015A&A...582A.112B}.

There are two main approaches to measure the disk mass. The first one is based on deriving gas mass from the dust continuum emission~{\citep{2011ARA&A..49...67W, 2005ApJ...631.1134A, 2023ASPC..534..501M}}, assuming a standard 100:1 ratio between gas and dust mass for the interstellar medium~\citep{1978ApJ...224..132B}, which is not necessarily applicable for disks~\citep{2016ApJ...828...46A}. The second, more direct method, determines the protoplanetary disk mass from observations of the emission lines of various molecules assuming a certain abundance ratio with respect to molecular hydrogen. In this respect, a suitable candidate is deuterated hydrogen HD~\citep{2013Natur.493..644B, 2016ApJ...831..167M, 2017A&A...605A..69T, 2020A&A...634A..88K}, which is well mixed with gas and has a fairly high abundance of $3\times10^{-5}$ relative to the molecular hydrogen~\citep{2013Natur.493..644B}. However, there are currently no instruments that could possibly observe the necessary HD lines after the end of \textit{Herschel} mission.

Having lost the option to measure HD, attention of the community turned to CO as possible gas mass tracer. \cite{2014ApJ...788...59W} (hereafter WB14) investigated CO isotopologue emission for a set of models with many varying parameters but with the following assumptions: a constant [CO]/[H$_2$]$= 10^{-4}$ ratio, complete freeze-out of CO onto dust at $T < 20$\,K, and complete photodissociation of CO where the surface density $N_{\rm H_2} < 1.3\times10^{21}$\,cm$^{-2}$. According to their results, the fluxes in the $^{13}$CO and C$^{18}$O lines allow us to estimate the disk mass within an order of magnitude. \citet{2016A&A...594A..85M} (hereafter M16) investigated a similar problem, but with more detailed chemistry, including isotope-selective reactions. However, results of both works lead to an underestimation of the disk mass by up to two orders of magnitude compared to the estimates from the dust and HD~\citep{2023ARA&A..61..287O}. Presumably, a more detailed account of CO chemistry, especially surface chemistry, and the inclusion of dust evolution in the disk would resolve the disagreement between the estimates. 

{\citet{2018A&A...618A.182B} followed CO chemical conversion on the grain surfaces in the environments of cold dense midplanes of T~Tauri and Herbig Ae/Be disks. They found that in colder T~Tauri stars, most of CO is indeed chemically processed there into other species on the Myr timescale and that not accounting for it can lead to the mass underestimation by up to two orders of magnitude.} \citet{2022ApJ...925...49R} included surface chemistry and argued that C$^{18}$O is a good tracer of mass, possibly even better than HD. They found this, however, with a very limited parameter survey. Obtaining the disk mass by their method also requires prior knowledge of disk dust mass that must be inferred from continuum observations. A method by~\citet{2022ApJ...926L...2T}, that uses a combination of C$^{18}$O and N$_2$H$^+$ to measure gas mass, also relies upon prior knowledge of disk structure from SEDs and resolved line and continuum observations. 

\cite{2017ApJ...849..130M} {represent a different approach to the problem: {instead of studying how to use CO as a mass tracer,} they searched for a gas mass tracer alternative to CO. They employed a  chemical model with gas-phase, ion, photo- and grain surface (but no isotopologue-selective) chemistry to look at abundance variations of a large list of molecules on a grid of model parameters.} CO abundance proved to be one of the most indifferent to changes in model parameters {and the authors concluded that it remains the best option for tracing disk mass. However, they only considered the species abundances, and did not conduct radiative transfer modeling.} 

In this work we conduct {a parameter survey simulating the molecular flux of CO isotopologues for protoplanetary disks around T~Tauri-type stars. We follow the approach of~\citetalias{2014ApJ...788...59W} and~\citetalias{2016A&A...594A..85M}, but use a model with grain surface chemical network~\citep[although without the isotope-selective processes,][]{2017ApJ...849..130M}. The use of the network with grain surface chemistry allows to account for the chemical depletion of CO from the gas phase~\citep{2018A&A...618A.182B}. Radiative transfer modeling is necessary to account for optical depth effects: line fluxes may be more sensitive to disk parameters than CO abundances, challenging the findings of~\citet{2017ApJ...849..130M}.} 

{The paper is structured as follows.} In Section~\ref{sec:model} we describe the models we use and introduce all our parameters. In Section~\ref{sec:res} we present our results, namely found dependences of flux on parameters, their approximation and physical explanation. {In Section~\ref{subsec:incl} we first go through the results and analysis related to the inclination to justify the usage of only one inclination ($i=60$\textdegree) further on in the work. In Section~\ref{sec:discussion}} we {discuss how our results} compare with~\citetalias{2014ApJ...788...59W} and~\citetalias{2016A&A...594A..85M} {and provide results from additional models to isolate effects of differences between our and their models}. Finally, in Section~\ref{sec:concl} we summarize our results.

\section{Model description}\label{sec:model}

We use the two-dimensional axisymmetric astrochemical model ANDES~\citep{2013ApJ...766....8A} with modifications {presented in}~\cite{2024MNRAS.527.7652Z} to calculate the physical and chemical structure of protoplanetary disks. The ANDES output is then transferred to the RADMC-3D radiative transfer code~\citep{2012ascl.soft02015D} using the DiskCheF\footnote{\url{https://gitlab.com/SmirnGreg/diskchef}} package to calculate the continuum emission and the line fluxes.

\subsection{ANDES}\label{subsec:andes}

The physical structure of the disk in the ANDES model is set up via the gas density, the dust temperature, and the UV radiation field. The disk is assumed to be axisymmetric, and the vertical structure is calculated assuming hydrostatic equilibrium. The radial structure is defined by a parametric surface density:
\begin{equation}
    \Sigma(R) = \Sigma_0 \left( \frac{R}{R_{\rm c}} \right)^{-\gamma} e^{-(R/R_{\rm c})^{2-\gamma}} e^{-(R_{\rm c,in}/R)^{2-\gamma}},
\end{equation}
where $\Sigma_0$ is a normalization constant obtained from a boundary condition, $\gamma$ parameterizes the radial dependence of the disk viscosity and $R_{\rm c}$ and $R_{\rm c,in}$ are outer and inner characteristic disk radius, respectively. The boundary condition for surface density is:
\begin{equation}
    {M_{\rm d} = 2 \pi \int_{R_{\rm min}}^{R_{\rm max}} \Sigma (R)\,R dR,}
\end{equation}
where $M_{\rm d}$ is the disk mass {and $R_{\rm min}$ and $R_{\rm max}$ are radial limits of the model spatial grid. For all models $R_{\rm min} = 1\,$au and $R_{\rm max} = 3000$\,au}.

The temperature in the disk atmosphere is calculated from the radiative transfer equation, taking into account radiative heating from the star, the accretion region, and the interstellar radiation field. The temperature in the disk midplane is determined separately, from the luminosities of the star and the accretion region~\citep{2017ApJ...849..130M}. The gas and dust temperatures are assumed to be equal. The luminosity $L_\star$ and radius $R_\star$ of the star are determined from its mass $M_\star$ using evolutionary models by~\citet{2008ASPC..387..189Y}, assuming an age of 1\,Myr. The luminosity of the accretion region is calculated from the relation $L_{\rm acc} = \dfrac{3}{2}\dot{M}\dfrac{GM_\star}{R_\star}$, where $\dot{M}$ is the accretion rate. We assume $\dot{M} = 10^{-8}\,M_\odot$ yr$^{-1}$. Accretion region size is obtained using the Stefan-Boltzmann law $R_{\rm acc} = \sqrt{\dfrac{L_{\rm acc}}{4\pi \sigma T_{\rm eff}^4}}$, where $T_{\rm eff} = 10^4$\,K. The calculation is performed up to a time moment of $10^6$ yr.

The dust ensemble is described by a power-law size distribution $dN\propto a^{-3.5}da$ with the exponent corresponding to the classical model of~\citep{1977ApJ...217..425M} for the interstellar matter (MRN model). The range of dust sizes is from $0.005$ $\mu$m to $25$ $\mu$m. The upper limit is increased compared to the MRN to account for dust growth to larger sizes than in the ISM. {The dust mass is taken equal to 0.01 of the gas mass. At each point of the disk, the dust size distribution is assumed to be the same; dust settling to the disk midplane and radial drift are not considered. Protoplanetary disks are evolving objects, and dust mass and size distribution should inevitably change during their lifetime \citep[see, e.g.][]{2014prpl.conf..339T}. The lack of consideration of dust evolution in our model can have an effect on distribution and emission of CO and its isotopologues, but we leave their detailed analysis for a later study.}

{Chemical composition of gas and ice in the ANDES model is calculated using the chemical reaction network from the ALCHEMIC model~\citep{2011ApJS..196...25S}. The network contains 650 species, and was updated by \citet{2017ApJ...849..130M} based on newer reaction rates from KIDA network \citep{2015ApJS..217...20W}. \citet{2024MNRAS.527.7652Z} updated binding energies of key species following \citet{2017SSRv..212....1C} and added the restriction of surface reaction rates to parameterize suppression of surface chemistry in deeper layers of the ice mantles below the three upper monolayers. The modified network includes 7807 reactions of the following types:}
two-body reactions in the gas phase, adsorption onto dust, desorption from dust, reactions on dust surface, photoreactions, reactions with cosmic rays and cosmic ray induced photons, and recombination reactions, including with dust grains. The probability of reactive desorption is 1\%. The ratio of diffusion energy to desorption energy is 0.5. Tunneling through reaction barriers is not taken into account for all species except H and~H$_2$.

The cosmic ray ionization rate
{is calculated using the prescription from \citet{2018A&A...614A.111P} (see their Appendix F, model \( \mathscr{H} \)).}
{The ionization rate also includes ionization by X-ray photons from the central source, which is calculated following \citet{2009ApJS..183..179B} for the X-ray luminosity of $10^{30}$\,erg~s$^{-1}$.}

{CO molecule participates in both gas-phase and surface reactions. One of the main parameters affecting CO abundance in the gas are freeze-out and desorption and photodissociation by UV photons (self-shielding of CO included), as accounted for by e.g.~\citetalias{2014ApJ...788...59W}. However, at longer timescales ($\sim$Myr) CO abundance is also affected by dissociation with ions (He$^+$) in the gas-phase, as well as surface reactions with H, O and OH \citep{2018A&A...618A.182B}. The model includes grain surface reactions beyond hydrogenation, which affect the abundance of CO in the ice and consequently in the gas phase. Particularly, the surface reaction with OH forming CO$_2$ leads to efficient CO depletion \citep{2018A&A...618A.182B}.}

It is important to note that there are no isotopologues in the chemical reaction network in ANDES, so the abundances of CO isotopologues are calculated from the abundance of the main isotopologue assuming the following ratios: $n(^{13}$CO$) = n($CO$) / 70$, $n($C$^{18}$O$) = n($CO$) / 550$~\citepalias{2014ApJ...788...59W}. Here and throughout the paper by C and O without superscripts we mean the main isotopes $^{12}$C and $^{16}$O. {However, to assess the effect of the isotope-selective processes, we do a parametric test using the isotope-selective shielding coefficients for $^{13}$CO and C$^{18}$O in Section~\ref{sec:discussion}.}

As initial abundances of chemical species, we take the chemical composition of a 1~Myr old molecular cloud, pre-calculated with temperature (10 K) and density ($10^{4}$ cm$^{-3}$) typical for molecular clouds and zero radiation field~\citep{2018ApJ...866...46M}. {To initialize the model, the composition is set to be the same everywhere in the disk.} {It is not entirely clear whether the initial composition of the protoplanetary disk should be atomic or molecular, i.e. inherited from the parent cloud \citep[e.g.,][]{2016A&A...595A..83E}, and many previous studies adopted atomic initial compositions \citep[e.g.,][]{2014A&A...572A..96M,2016A&A...594A..85M,2022ApJ...925...49R}. More recent observations and models, particularly the detection of methanol in protoplanetary disks, indicate that a lot of ices must be preserved to reproduce the observational data, thus favoring the inheritance scenario \citep{2021NatAs...5..684B,2025arXiv250204957E}. Following these findings, we assume inheritance scenario and adopt molecular initial chemical composition.}

The main differences of our model compared to the previous parameter surveys are inclusion of grain surface chemistry, using molecular initial conditions and lack of isotope selective processes. {Further on, we will refer to our chemical network as grain surface chemistry network.}

\subsection{RADMC-3D}\label{subsec:radmc}

To simulate molecular line emission together with the continuum, we use the radiative transfer code RADMC-3D~{\citep{2012ascl.soft02015D}}. As input, we provide CO number density, dust density and temperature distributions from ANDES. We also supply RADMC with inclination of the disk $i$ (0\textdegree\ is face-on), Keplerian velocity field throughout the whole disk and molecular line data from the LAMDA database~\citep{2005A&A...432..369S}. For all three isotopologues we choose the $J=2-1$ transition as one of the most widely used in observations~\citep{2024MNRAS.527.7652Z}. Energy level populations are calculated assuming LTE. For each line we use a spectral window of 40\,km s$^{-1}$ with 0.2\,km s$^{-1}$ resolution centered at the rest frequency of the line. The distance to every model disk is set to be 150\,pc.

As output, we get spectral cubes which we subtract the continuum from and then integrate over solid angle and velocity to obtain the total flux from the disk. Continuum level is determined as the median of emission in channels with velocity beyond $\pm15\,$km s$^{-1}$.

\subsection{Set of models}

As in previous works~\citep{2014ApJ...788...59W, 2016A&A...594A..85M, 2017ApJ...849..130M}, we create a grid of models with varying disk parameters to study the effect they have on molecular emission. Out of all introduced parameters, we choose to vary the star mass $M_\star$, the disk mass $M_{\rm d}$, the outer characteristic disk radius $R_{\rm c}$, and the inclination of the disk $i$ as main influences of disk structure and emission. Further on, $R_{\rm c}$ will be referred to as just characteristic disk radius. We vary them within ranges typical of T~Tauri stars \citep{2011ARA&A..49...67W,2023ASPC..534..501M}. The values that the parameters take are summarized in Table ~\ref{tab:models}. In total, there are 1280 models in the set.

{The parameter $\gamma$ determines the radial density profile of the disk, from below zero \citep{2017A&A...606A..88T}, to around~$1$ \citep[e.g.,][]{2020A&A...634A..88K,2021A&A...646A...3L} and $1.5$ in the classic minimum mass solar nebula (MMSN) model \citep{1977Ap&SS..51..153W}. Here we do not intend to explore the effect of this parameter, as it should be essentially similar to that of $R_{\rm c}$, determining the fraction of mass bound in the inner {disk} regions, and adopt a typical value of $\gamma=1$.} 
Inner characteristic radius $R_{\rm c,in}$ is not varied as it would mainly influence the inner disk that the low-J CO does not trace. In all models $R_{\rm c,in}=3\,$au{, while the inner boundary of the spatial grid is 1\,au, which provides a smooth density change in the inner disk. This allows to avoid high density and temperature at the inner cells of the computational domain, which are directly illuminated by the central source due to the absence of grid points closer to the star.}

\begin{table}[h!]
    \centering
    \caption{Values of parameters in the set}
    \begin{tabular}{cc}
\hline \hline
Parameter & Values \\ \hline
$M_\star$ & 0.5, 1.0, 1.5, 2.0\,$M_\odot$ \\\hline
\multirow{2}{\widthof{$M_{\rm d}$}}{$M_{\rm d}$} & 0.001, 0.002, 0.005, 0.01,  \\
 & 0.02, 0.05, 0.1, 0.2\,$M_\odot$\\\hline
$R_{\rm c}$ & 12.5, 25, 50, 100, 200\,au \\\hline
\multirow{2}{\widthof{$i$}}{$i$} & 0\textdegree, 30\textdegree, 45\textdegree, 60\textdegree, \\ 
 & 75\textdegree, 80\textdegree, 85\textdegree, 90\textdegree\\\hline
    \end{tabular}
    \label{tab:models}
\end{table}

\section{Results}\label{sec:res}

\subsection{{CO depletion and its snowline}}
\label{sec:CO_depletion}

\begin{figure*}
    \centering
    \includegraphics[width=\textwidth]{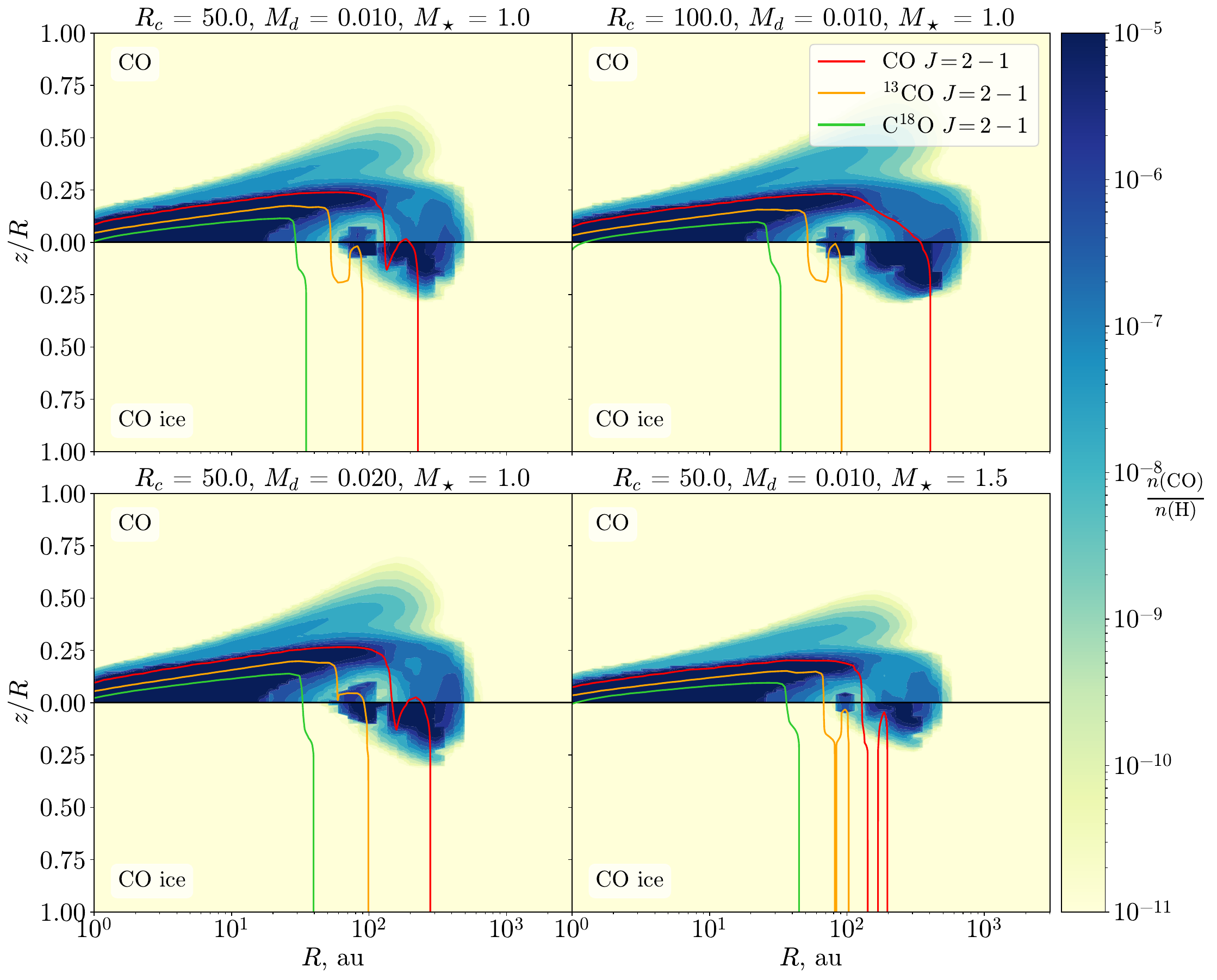}
    \caption{{CO gas and ice abundance distributions in the disk for four models with different model parameters. Contours represent $\tau=1$ surfaces of used lines. Left upper panel: a reference model with $M_{\rm d} = 0.01\,M_\odot$, $R_{\rm c} = 50$\,au, $M_\star=1.0\,M_\odot$. Other panels represent models with one parameter value different from the reference. Right upper panel: $R_{\rm c} = 100$\,au. Left lower panel: $M_{\rm d} = 0.02\,M_\odot$. Right lower panel: $M_\star=1.5\,M_\odot$.}}
    \label{fig:tau1}
\end{figure*}

{One of the main processes affecting the CO abundance in the disk is CO depletion. Gas-phase CO can be depleted by three main processes: freeze-out to dust grain surface, photo-dissociation, and chemical depletion, i.e. transformation to other ice- or gas-phase species. The first two processes can be accounted for parametrically, by assuming that CO gas is absent below a certain temperature and at low column densities, respectively, as implemented e.g. by \citetalias{2014ApJ...788...59W}. Chemical depletion requires more detailed consideration of surface chemistry and was suggested as a possible explanation of low observed CO column densities \citep{2013ApJ...776L..38F,2016ApJ...819L...7N} and investigated in multiple astrochemical simulations of protoplanetary disks \citep{2016A&A...595A..83E,2017ApJ...849..130M,2018A&A...618A.182B}.}

{Depletion of CO in the disks of T~Tauri stars was studied using astrochemical modeling by \citet{2018A&A...618A.182B}. Three main pathways are gas-phase destruction by He$^{+}$ (active at longer timescales), grain surface conversion of CO to CO$_2$ inside the CO snowline, and the formation of CH$_3$OH in ice mantles outside the CO snowline. The most efficient reaction is the conversion of CO to CO$_2$ ice inside the CO snowline, which happens to the CO molecules temporarily accreted to dust surface. This process is also accounted for and active in our simulations, leading to low CO abundance in both gas and ice phase in a large area in the disk midplane. In our modelling, it is also more efficient and covers a broader range of temperatures than 20-30\,K, which is in agreement with the predictions of \citet{2018A&A...618A.182B} of the efficiency of this process at different dust sizes and CO binding energy. According to their Appendix~B.4.3 and B.4.5, increase in dust size and the CO binding energy should lead to higher efficiency of this process. The average dust size used for chemistry in our model is $3.5\cdot10^{-3}$\,cm (cf. their $10^{-5}$\,cm), and the CO binding energy we use is 1180\,K, (cf. their 855\,K). Other authors found the effect of the chemical depletion at 1\,Myr timescale to be less significant than in our results \citep[e.g.][]{2019ApJ...883...98Z,2020ApJ...899..134K}. The difference comes from the use of different chemical networks (UMIST vs. ALCHEMIC), as well as larger grain size and higher CO binding energy in our modelling.}

{Chemical depletion makes it difficult to determine the position of the CO snowline, because it interferes with the balance of adsorption and desorption that is usually considered to define the snowlines \citep[e.g.][]{2015A&A...582A..41H}. The distribution of CO in the gas and in the ice for selected models is shown in Fig.~\ref{fig:tau1}. As can be seen from {Fig.}~\ref{fig:tau1}, at a range of distances CO is not abundant in either the gas or the ice. In these models, gas-phase CO is abundant in the midplane inside $\sim20$\,au; beyond that and up to the distance of $60-80$\,au it is absent in both phases; ice-phase~CO appears at larger distances $>60-80$\,au. The particular distances where these transitions happen vary across the model, and so do the temperatures at these locations. Chemical depletion creates this intermediate region of low CO abundance in both phases, which call for an alternative definition of the snowline position.} 

{We can define the CO snowline as the position in the disk midplane where ice- and gas-phase abundances of CO become equal. This will be the location determined by the freeze-out, and the corresponding midplane temperature will be referred to as the freeze-out temperature $T_{\rm freeze}$. Additionally, we can find a location in the disk midplane where gas-phase CO abundance falls below certain value ($10^{-5}$ w.r.t. H~nuclei), which should happen closer to the star. This location will be characterized by the chemical depletion temperature $T_{\rm chem}$, which is higher than $T_{\rm freeze}$. Thus, at $T > T_{\rm chem}$ CO is abundant in the gas, at $T < T_{\rm freeze}$, CO is abundant in the ice, and between these temperatures, it is depleted from both phases.}

\begin{figure}
    \centering
    \includegraphics[width=0.5\textwidth]{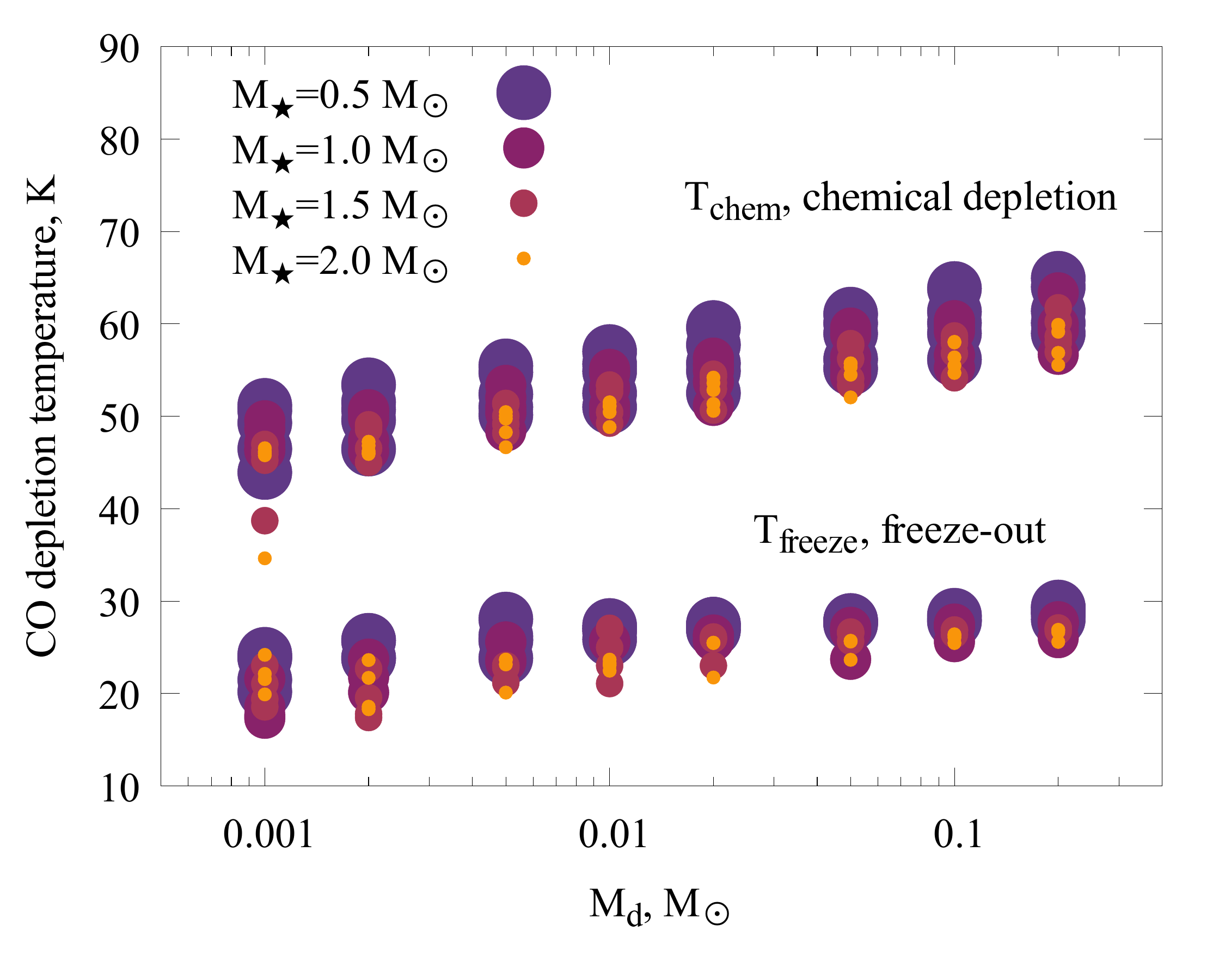}
    \caption{{Freeze-out and chemical depletion temperatures depending on the disk mass.}}
    \label{fig:Tfreezechem}
\end{figure}

{We calculate both $T_{\rm chem}$ and $T_{\rm freeze}$ for each model and show them in {Fig.~\ref{fig:Tfreezechem}} depending on disk mass. For the smallest considered disks ($R_{\rm c}=12.5$\,au), there is no $T_{\rm freeze}$, because the temperature never drops low enough inside the disk, while outside the disk CO ice is photodesorbed. {Fig.~\ref{fig:Tfreezechem}} shows that both freeze-out and chemical depletion temperatures vary among the models. They are typically higher in more massive models due to higher density in the midplane. For both freeze-out and chemical depletion, this results from the higher collision rate of CO molecules with dust grains. Increasing stellar mass has an opposite effect, because in disks around more luminous stars, similar temperatures are reached at larger distances, where densities are lower. Overall, there is a wide range of CO {chemical} depletion temperatures: from $34-51$\,K in the $0.001$\,$M_{\odot}$ disks, to $55-65$\,K in the {$0.2$}\,$M_{\odot}$ disks. Freeze-out temperatures of CO also vary between 17~and 29\,K.}

\subsection{{Dependence of CO isotopologue emission on disk parameters}}

The variation of the model parameters has evident influence on the physical and chemical structure of the disk, in particular on the abundance of CO isotopologues, and hence on their line emission. In this Section, the impact of each individual parameter on the physics and chemistry of the disk is further discussed in detail.

 {In Fig.~\ref{fig:tau1},} CO gas and ice abundance distributions and $\tau=1$ isolines for the considered molecular transitions are illustrated for four different models of our grid. One of the models is the `reference model' with average parameters in our {grid} ($M_{\rm d} = 0.01\,M_\odot$, $R_{\rm c} = 50$\,au, $M_\star=1.0\,M_\odot$), and the other three models explore the variation in the disk mass, size and the mass of the central star. We calculate the optical depth in the lines of CO and its isotopologues assuming the face-on disk observation. The $\tau=1$ isoline of CO and $^{13}$CO shift quite significantly due to the complex distribution of gas-phase CO in the molecular layer and the midplane, while the $\tau=1$ for C$^{18}$O appears to be less sensitive to the variation of the disk model parameters.

\begin{figure*}
    \centering
    \includegraphics[width=1\linewidth]{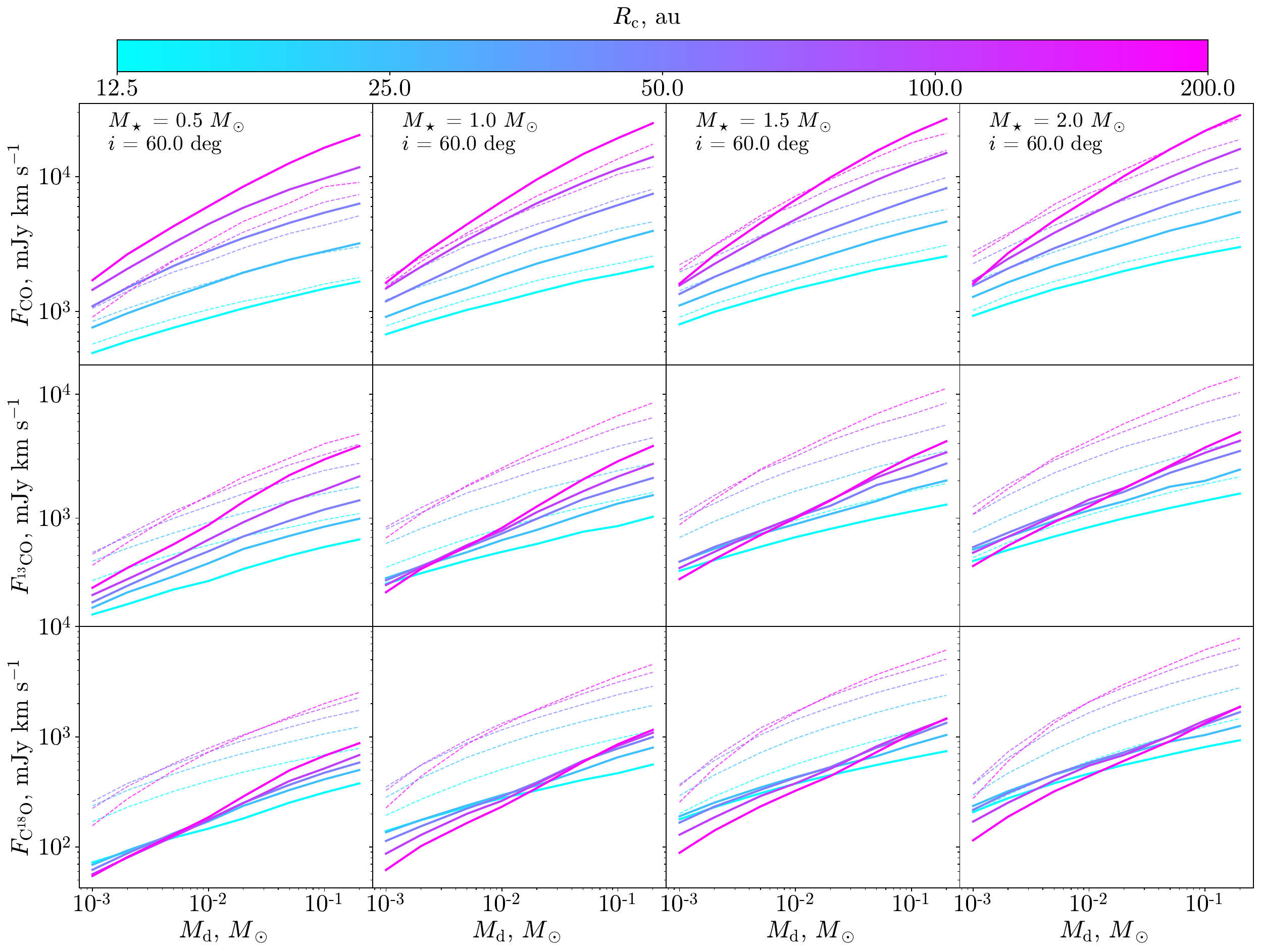}
    \caption{Dependence of CO, $^{13}$CO and C$^{18}$O line fluxes on disk mass, disk radius and stellar mass at the disk inclination $i=60$\textdegree. {Solid lines are results with {grain surface chemistry} network while dashed lines are results with assumptions from~\citetalias{2014ApJ...788...59W}.}}
    \label{fig:mdi60}
\end{figure*}

{Fig.~\ref{fig:mdi60} shows the dependence of the line fluxes on the system parameters: disk mass $M_{\rm d}$, characteristic disk radius $R_{\rm c}$, and star mass $M_\star$ at a constant disk inclination $i=60$\textdegree. Solid lines are for our grain surface chemistry network.} Dashed lines show the same dependences in the models without chemistry, where we follow~\citetalias{2014ApJ...788...59W} and calculate the abundance of CO parametrically, assuming it is only restricted by freeze-out at 20\,K and photodissociation. For the rest of the Section~{\ref{sec:res}}, we focus on {the grain surface chemistry} case only. We will return to the comparison with the results of other authors in Section~\ref{sec:discussion}.

\subsubsection{{Disk inclination $i$}}\label{subsec:incl}

\begin{figure*}
    \centering
    \includegraphics[width=1\linewidth]{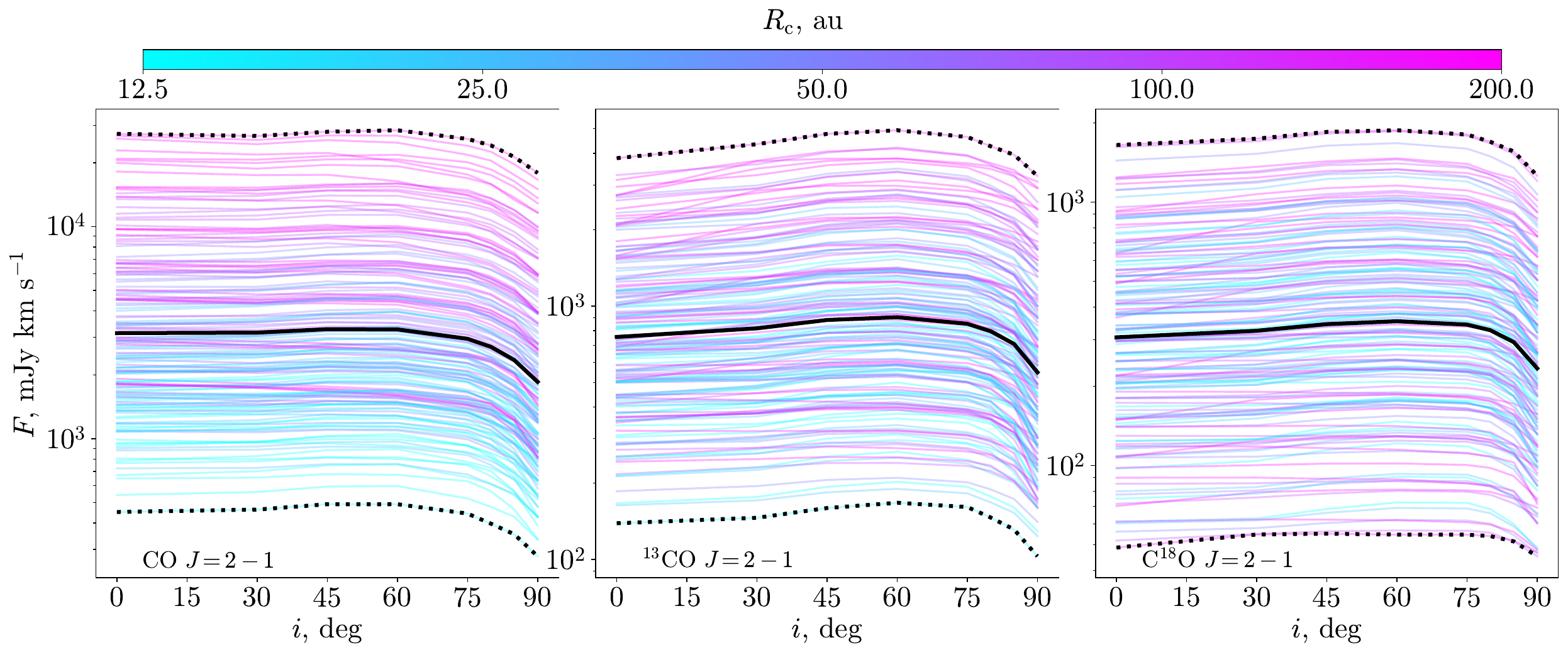}
    \caption{Dependence of CO, $^{13}$CO, and C$^{18}$O line fluxes on disk inclination. Along the individual colored lines, all parameters except the inclination are fixed. The black dashed lines show the maximum and minimum flux values at each inclination. Black solid lines show geometric mean of flux dependences.}
    \label{fig:flux_i}
\end{figure*}

Fig.~\ref{fig:flux_i} demonstrates the dependence of the isotopologue line fluxes on the disk inclination for each set of fixed parameter values, as well as the maximum, minimum, and geometric mean of dependences at each inclination. For CO, on average, the flux is independent of the inclination up to $i=60$\textdegree, while at larger angles the flux drops slightly {(less than a factor of two)}. For C$^{18}$O and $^{13}$CO there is a slight rise {($1.2F(i=0$\textdegree) at peak)} of flux from 0\textdegree\ to 60\textdegree\ and then a similar drop. Due to the found weak dependence and the lack of qualitative influence on the dependences on the other parameters, only the case of the most probable inclination $i=60$\textdegree\ will be considered further.

The dependence can be explained the following way. On the one hand, increasing the inclination of the disk increases the amount of matter in the line of sight while decreasing the total visible area of the disk. In the case of CO, it leads to a hotter but smaller $\tau=1$ disk surface and appearance of colder disk outer edge. In the case of C$^{18}$O and $^{13}$CO, while we also see less of the disk surface, $\tau=1$ surface at low $i$ can increase in size due to more matter in the line of sight in previously transparent areas, thus flux can slightly increase as well. For higher inclinations, the disk stops being transparent anywhere and flux decreases as in the case of CO. 

On the other hand, the Doppler effect becomes more prominent as the projection of the rotational velocity component of the disk onto the line of sight increases. This redistributes the energy across the spectrum and reduces the optical depth at the central line frequencies, allowing us to see more matter overall. As a result, the flux increases. The combination of these two effects keeps the resulting flux constant at $i<60$\textdegree\ for CO and slightly rising for C$^{18}$O and $^{13}$CO. At $i>60$\textdegree\, the first effect becomes more significant and the flux drops.

\subsubsection{Characteristic disk radius $R_{\rm c}$}\label{subsubsec:radius}

Increasing the disk radius at a fixed disk mass {({cf.} upper panels of Fig.~\ref{fig:tau1})} leads to redistribution of mass from the inner to the outer parts of the disk. The outer parts are colder, and at some distance from the star {CO gets depleted from the gas phase due to chemical depletion and freeze-out. Beyond a certain distance, where the temperature is lower than $T_{\rm chem}$ shown in {Fig.~\ref{fig:Tfreezechem}}, gas-phase CO abundance becomes low and it does not emit.} As a result, larger disks, on the one hand, can have more flux due to more matter emitting from a larger disk surface, but on the other hand, it can decrease due to the depletion of more CO.

Which of the factors will dominate depends on the optical depth of the line emission. If the line emission is optically thick, then a lot of matter is under the $\tau=1$ surface and does not contribute to the total flux. Increasing the size of the disk brings some of this previously invisible matter into new regions and it becomes visible, which increases the flux. However, this redistribution shifts the $\tau=1$ surface closer to the midplane where temperatures are lower, so the flux from the inner parts decreases a bit. This is the case of CO {flux, which always increases with radius in the upper panels of Fig.~\ref{fig:mdi60}}. The emission in this line is so optically thick that the $\tau=1$ surface is almost coincident with the upper boundary of the molecular layer. Freezing has no effect on the flux in this line, since it occurs near the disk midplane, under the $\tau=1$ surface. 

If the line emission is optically thin, then the redistribution of matter reduces the flux, since initially, emission from almost all the matter is already observable, and with the disk size increased, the matter is redistributed to colder regions. The freeze-out also contributes to the flux decrease. This case is relevant to C$^{18}$O in low-mass disks. {In lower panels of Fig.~\ref{fig:mdi60} its flux can be seen first falling with radius for low-mass disks and increasing after some disk mass value that depends on the stellar mass.} The disk, particularly the inner disk, is not completely transparent in this line, but enlarging the disk lowers the optical depth enough to cause a noticeable loss of flux. At larger disk masses, however, the removal of matter from optically thick regions becomes more significant in its contribution to the flux. The case of $^{13}$CO is intermediate between CO and C$^{18}$O.

{Overall, in the optically thick case, radius increase leads to an increase of the line flux, while in the optically thin case, it causes a decrease in the line flux.}

\subsubsection{Disk mass $M_{\rm d}$}
    
An increase in the disk mass {(left panels of Fig.~\ref{fig:tau1})} leads to an increase in the absolute CO mass density in the entirety of the disk. This shifts the $\tau=1$ surface away from the star both radially and vertically, with the magnitude of this shift increasing with the radius of the disk $R_{\rm c}$. Temperature in the disk rises with height so flux increases from the entire $\tau=1$ surface which now is also larger. As a result, flux increases with disk mass in both optically thick and thin cases. {This is also reflected in Fig.~\ref{fig:mdi60} where all line fluxes rise with disk mass.} {The temperature range of the CO-free region in the midplane increases with mass (see {Fig.~\ref{fig:Tfreezechem}}, but inner edge of this region (where CO becomes more depleted) is hidden under the $\tau=1$ surface, and does not contribute much to the emission.}

\begin{figure*}
    \centering
    \includegraphics[width=\linewidth]{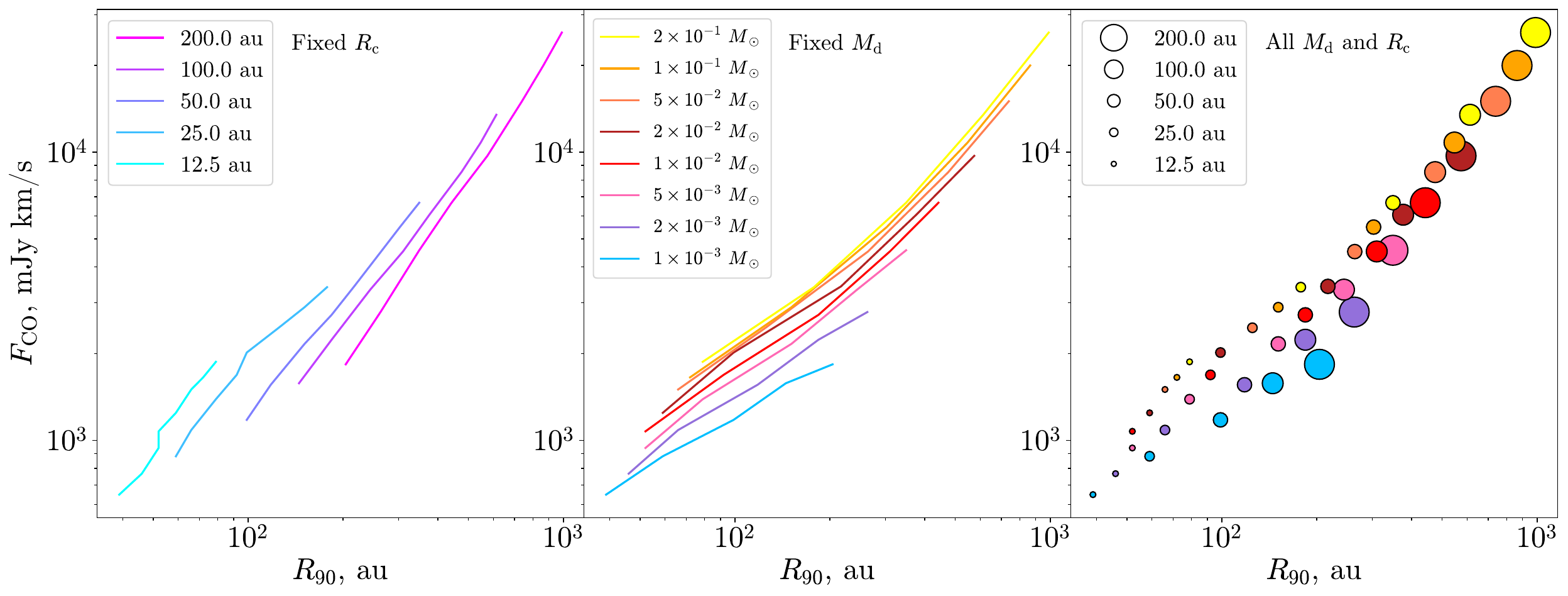}
    \caption{Dependences of the CO line fluxes on the $R_{90}$ at disk inclination $i=0$\textdegree\ and stellar mass $M_\star=1\,M_\odot$. Left panel presents dependences of fixed $R_{\rm c}$, middle presents of fixed $M_{\rm d}$ and right shows models of all $R_{\rm c}$ and $M_{\rm d}$ differentiated by shade and size.}
    \label{fig:r90}
\end{figure*}

The influence of disk mass on its size can be seen through another observational characteristic, $R_{90}$, which is defined as the radius of disk area that produces 90\% of the CO flux. In Fig.~\ref{fig:r90} we show how $R_{90}$ relates to the total CO flux $F_{\rm CO}$ for models of all chosen $M_{\rm d}$ and $R_{\rm c}$ at $i=0$\textdegree\ and $M_\star=1\,M_\odot$. On the left panel, we can see how both $R_{90}$ and $F_{\rm CO}$ increase with disk mass at fixed $R_{\rm c}$. Compared to the middle panel which shows the same but with $R_{\rm c}$ changing and $M_{\rm d}$ fixed, we see that {the slope of $F_{\rm CO}$-$R_{90}$ relation is different depending on which parameter is fixed}. This points at other factors that influence flux besides sheer size, and this factor is the increase in temperature at the $\tau=1$ surface as stated in the previous paragraph. Increase in $R_{\rm c}$ does not reflect a pure dependence of flux on size as well as it leads to a decrease in temperature at the $\tau=1$ surface (see Section~\ref{subsubsec:radius}). Hence, looking at a full picture on the right panel, we can conclude that the size of the disk cannot be accurately inferred from the CO flux alone as it requires a knowledge of thermal structure that is influenced by both $M_{\rm d}$ and $R_{\rm c}$.

\subsubsection{Stellar mass $M_\star$}

An increase in the mass of the star {(left upper and right lower panels of Fig.~\ref{fig:tau1})} leads to a stronger gravitational field and an increase in luminosity. The enhanced gravitational field compresses the disk in the vertical direction and increases the rotation velocity of the disk. The compression of the disk shifts the $\tau=1$ surface closer to {the} midplane. A higher rotational velocity gives a higher Doppler shift of the lines, which distributes the energy more broadly across the spectrum, allowing more matter to be seen, and hence the total flux increases. But this only has an effect on a model with an inclination other than zero. The effect is greater for inclinations closer to 90\textdegree.

The increase in luminosity warms the disk to a higher temperature and increases the {radiation field. It affects both CO emitting properties and local chemical reaction rates. In regard of the emission, higher} temperature increases the source function value $S_\nu$, but decreases the number density of CO emitting in the line, since the maximum $J=2$ level population of CO isotopologues corresponds to a temperature of $\sim$17\,K. This temperature is below a typical CO freeze-out temperature of 20\,K so it is impossible to see CO at such temperatures except at the very periphery of the disk (>1000\,au) where the photodesorption is active. These are counteracting factors, since the intensity $I_\nu \propto S_\nu (1 - e^{-\tau_\nu})$. Which factor is the dominant one depends on the disk radius, since the larger the disk is at a fixed disk mass, the more transparent it is. 

{Higher stellar luminosity means that CO depletion and freeze-out fronts move away from the star. The temperature of CO chemical depletion $T_{\rm chem}$ is lower in models with higher $M_{\star}$ (see {Fig.~\ref{fig:Tfreezechem}}), which means that the region of CO depletion between $T_{\rm chem}$ and $T_{\rm freeze}$ shifts even further out. On the one hand, the increased area of the region inside $T_{\rm chem}$ contributes to the flux increase. On the other hand, it is counteracted by CO destruction that becomes more efficient as the disk scale height decreases with stellar mass. As the disk shrinks to the midplane due to stellar gravity, the molecular layer moves closer to the midplane and becomes thinner (see lower panels of {Fig.}~\ref{fig:tau1}), it is more affected by the chemical depletion, effective at the midplane temperatures and densities. CO abundance in the molecular layer decreases due to the combined effect of chemical depletion and photodissociation, which is more significant in large low-mass disks, where the molecular layer above the depleted region basically disappears. This effect decreases the CO flux, particularly in low-mass disks more transparent to dissociative UV radiation. At the same time, the intrinsic UV radiation field does not change much with the stellar mass, as it is dominated by the emission from the hotter accretion region with the temperature of 10,000\,K.}

{Overall, stellar mass produces a range of effects with some working against each other in terms of flux. Which factors will overweigh the others depends on disk mass and radius, as well as the optical depth of the line in the CO depletion region.} {We discuss the effects of different contributions of stellar masses in more detail in Appendix~\ref{app:stellar_effects}.}

\subsection{Approximation of emission dependence on disk mass}\label{subsec:approx}

All the dependences in Fig.~\ref{fig:mdi60} can be approximately described by a linear regression in logarithmic scale of the following form
\begin{equation}
    \lg{f_l} = a_l + b_l\lg{m_{\rm d}},
\end{equation}
where $f_l = \dfrac{F_l}{\text{1\,mJy km/s}}$, $l$ is the line of any of the CO isotopologues and $m_{\rm d} = \dfrac{M_{\rm d}}{10^{-3}\,M_\odot}$. In this case, the approximation coefficients $a_l(R_{\rm c}, M_\star)$ and $b_l(R_{\rm c}, M_\star)$ will be functions of the remaining model parameters. This will allow us to analyze the dependences in more detail and compare them with each other both for different lines and at different fixed values of the remaining parameters. The approximation is performed using the routine \texttt{scipy.stats.linregress}\footnote{\url{https://docs.scipy.org/doc/scipy/reference/generated/scipy.stats.linregress.html}}.

\begin{figure*}
    \centering
    \includegraphics[width=0.92\linewidth]{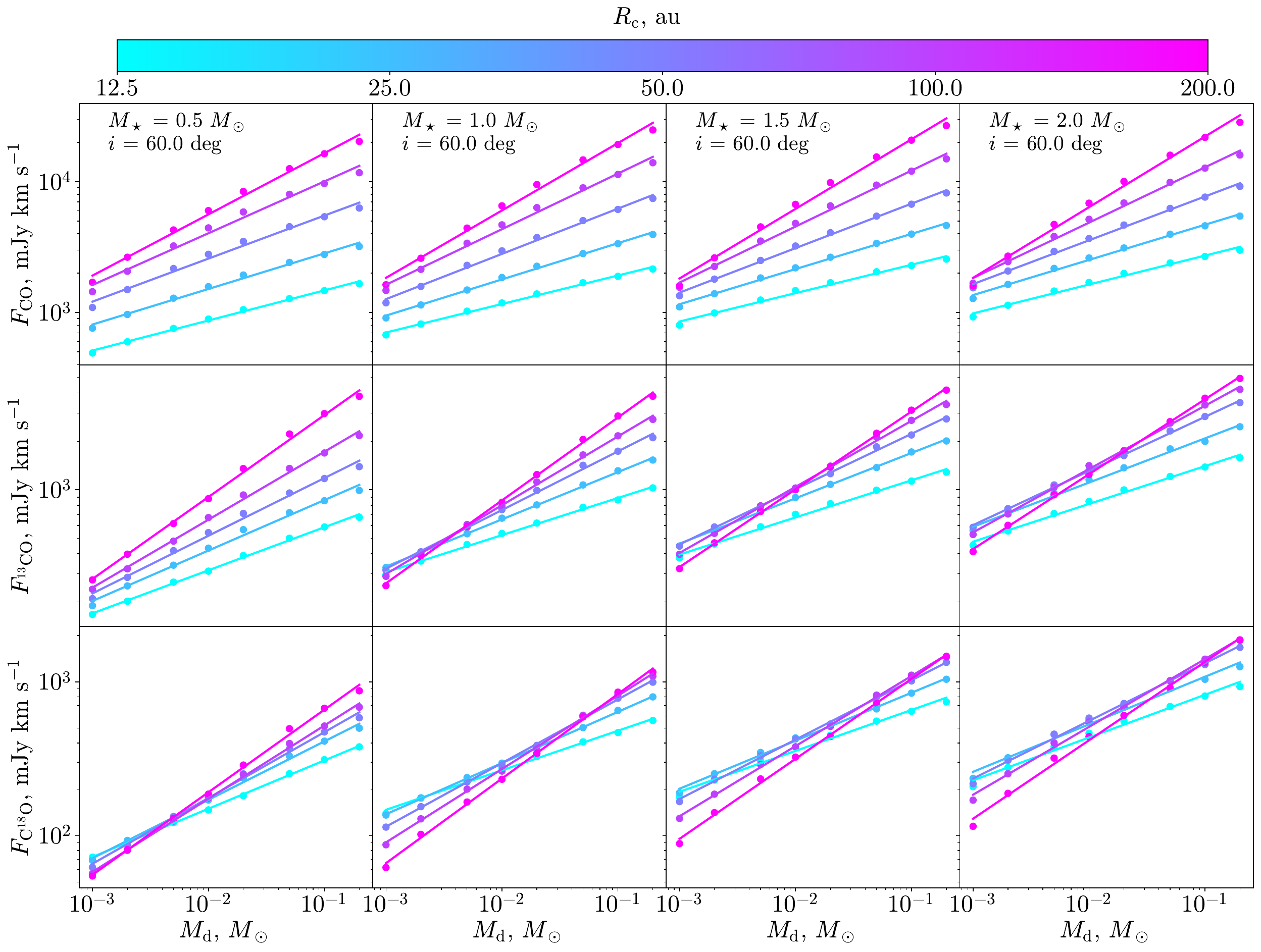}
    \caption{Dependences of the CO, $^{13}$CO, and C$^{18}$O line fluxes on the disk mass, disk radius, and stellar mass at disk inclination $i=60$\textdegree. The dots represent model values of the fluxes, the lines show the approximation.}
    \label{fig:mdi60fit}
\end{figure*}

\begin{figure*}
    \centering
    \includegraphics[width=\linewidth]{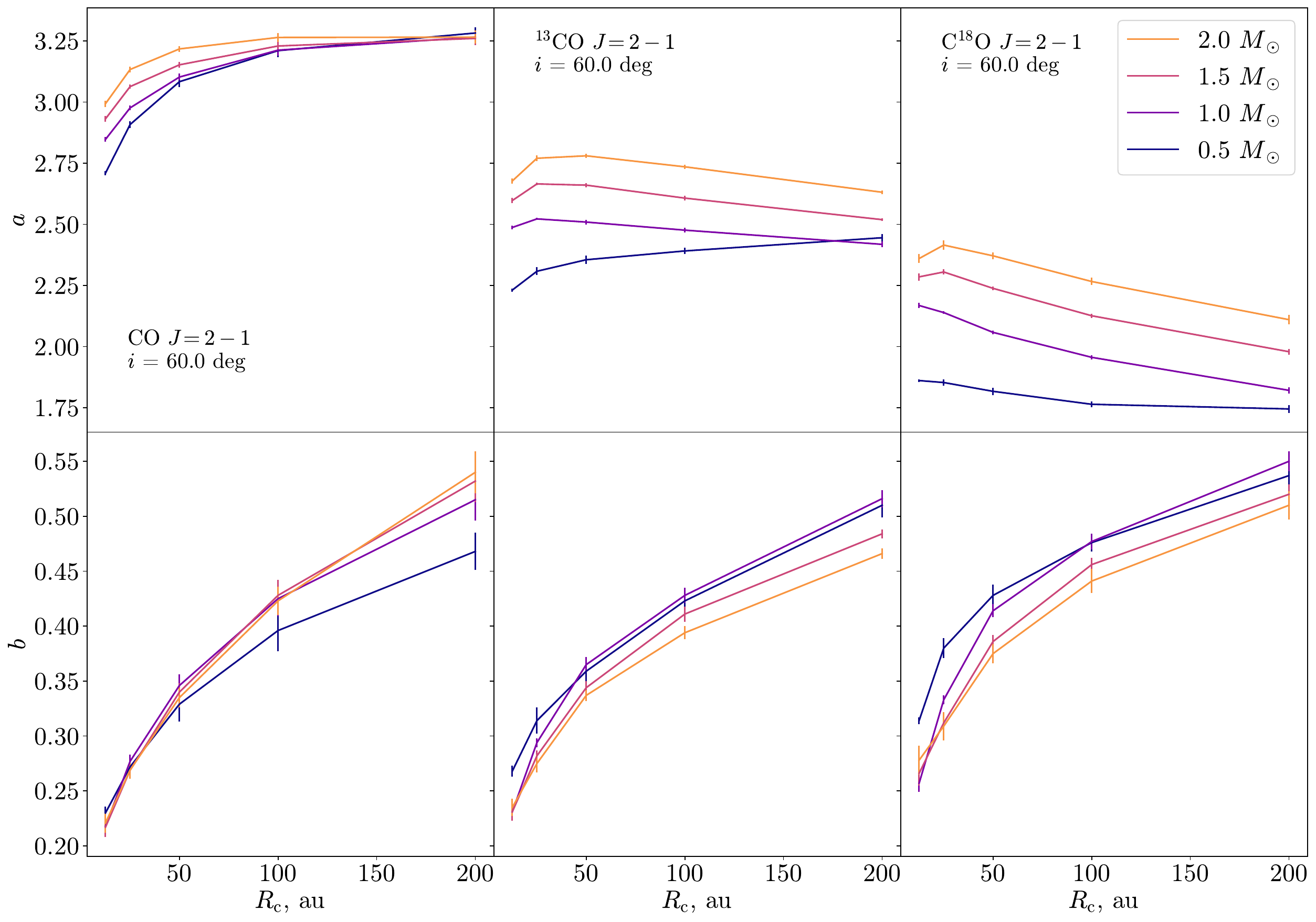}
    \caption{Dependences of the approximation coefficients on the disk radius and stellar mass at disk inclination $i=60$\textdegree\ for the CO, $^{13}$CO, and C$^{18}$O lines. The vertical bars denote the uncertainty of the approximation coefficient.}
    \label{fig:coefi60}
\end{figure*}

Fig.~\ref{fig:mdi60fit} shows the result of the approximation of the data from Fig.~\ref{fig:mdi60}, and Fig.~\ref{fig:coefi60} shows the dependence of the obtained coefficients $a_l$ and $b_l$ on the radius $R_{\rm c}$ and mass of the star $M_\star$. The intercept $a_l$ is the flux for models with $M_{\rm d} = 10^{-3}\,M_\odot$ by definition. For CO, $a_l$ rises with stellar mass for smallest disks, which are least affected by increasing {destruction in the molecular layer} and benefit more from the increased temperature. As the disk size increases, {chemical depletion and photodissociation in the molecular layer play} a bigger role so the flux decreases with stellar mass for the largest disks. For C$^{18}$O and $^{13}$CO, while the same factors come into play, the major contribution to flux comes from CO depletion area which is fully transparent in both lines. Therefore, when stellar mass is increased, more CO is introduced to the disk and the disk is overall hotter, leading to significantly higher molecular flux. Dependence on the disk radius is in full accordance with explanations in Section~\ref{subsubsec:radius}.

The slope $b_l$ reflects the sensitivity of the flux to the disk mass and hence the possibility to determine the disk mass from the flux. Fig.~\ref{fig:coefi60} shows that all lines are approximately equally sensitive to the disk mass. Additionally, the dependence on $R_{\rm c}$ does not change much with the stellar mass for CO line. The only exception for CO is presented by the largest disks which are most affected by {molecular layer destruction. This effect decreases with disk }mass and this leads to a change in slope. For C$^{18}$O and $^{13}$CO, there is a minor decrease in slope with increasing stellar mass as massive disks are slightly less affected by factors described in the previous paragraph, e.g. it is harder to heat a more massive disk, hence a smaller fraction of CO is desorbed. {Finally}, $b_l$ increases by a factor of 2-3 from the minimum radius to the maximum radius.

As a result, the fluxes in CO isotopologue lines can differ by an order of magnitude or more between models. Although flux dependences on individual parameters are clearly distinguished, the coefficients depend on other parameters with a remaining scatter of one order of magnitude in the flux. Hence, it is not possible to accurately determine the value of any parameter from the flux in a single CO line without the knowledge of the other parameters. It is possible that the simultaneous use of several lines could allow the determination of some parameters.

{It is important to note here a seeming discrepancy between the conclusion of an earlier study by~\citet{2017ApJ...849..130M} and our results. \citet{2017ApJ...849..130M} find that CO gas mass correlates well with total disk mass with correlation being stable against other parameters. As we use the same chemical model, CO gas mass in our work behaves the same way. But flux also depends on the structure and temperature profile of the emitting surface. They are much more sensitive to model parameters, hence the flux variations.}

\subsection{Using combinations of lines to determine disk parameters}\label{subsec:linecombo}

\begin{figure*}
    \centering
    \includegraphics[width=\linewidth]{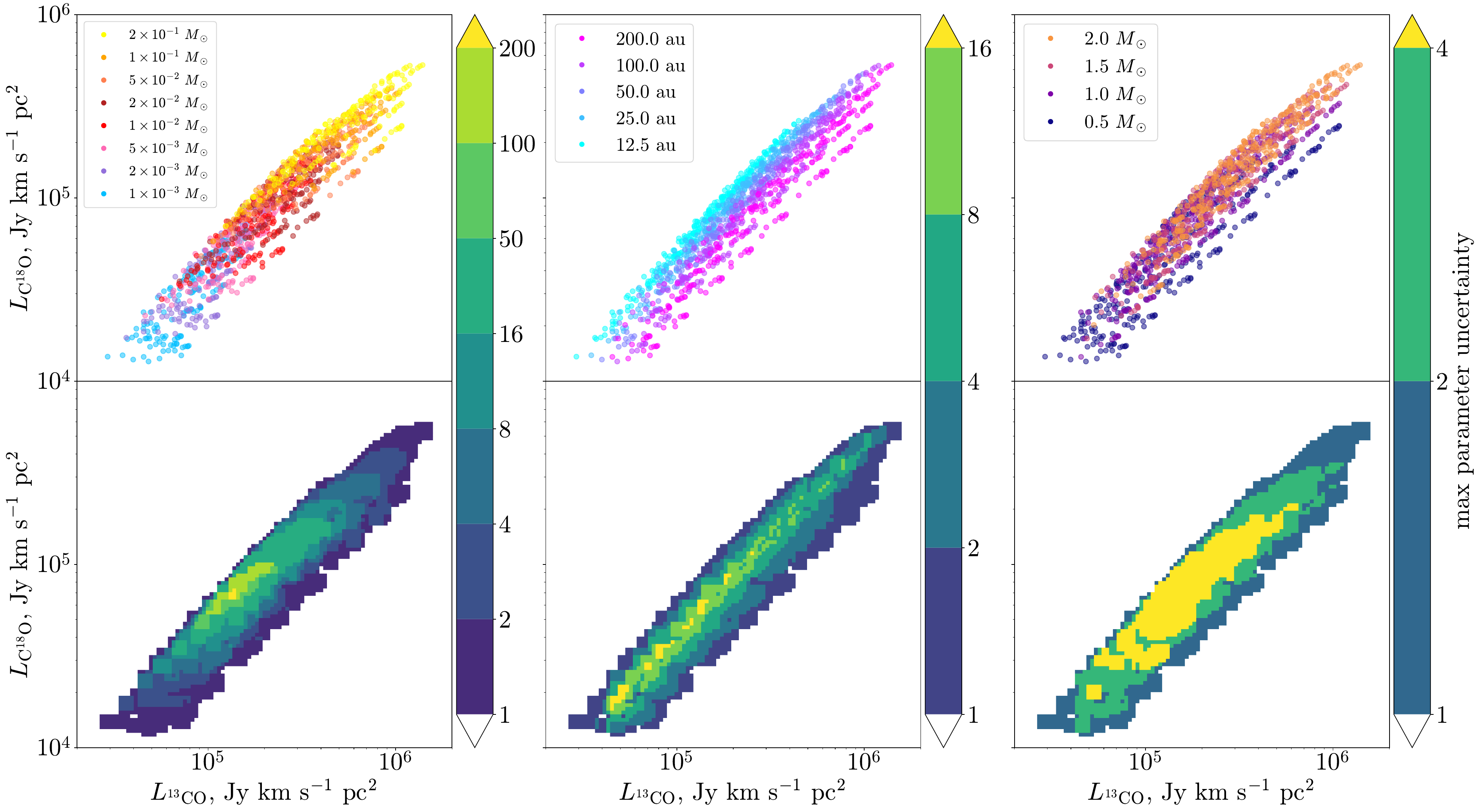}
    \caption{{Luminosity diagrams for the C$^{18}$O and $^{13}$CO lines in our models. Upper left: model fluxes are grouped by disk mass and contours outline the regions where models of corresponding mass reside. Upper middle: models are colored according to characteristic radius. Upper right: models are colored according to stellar mass. Lower left: distribution of uncertainty in disk mass. Lower middle: uncertainty in characteristic radius. Lower right: uncertainty in stellar mass. Uncertainty here is defined as the ratio of maximum and minimum parameter value within 10\% range of luminosities of the point on the diagram. White means there are no models nearby.}}
    \label{fig:lumlumus}
\end{figure*}

The use of two CO lines to determine parameters, in particular disk masses, was previously proposed in~\citetalias{2014ApJ...788...59W}. The authors considered a set of disk models with different parameter values and modeled CO isotopologue line emission from them. They plotted the resulting line luminosities ($L_l = 4\pi d^2 F_l$, where $d$ is the distance to the object) for C$^{18}$O and $^{13}$CO on a luminosity diagram {grouping them by disk mass} (Fig.~6 of their work or Fig.~\ref{fig:lumlumWB14} of this work). {We adopt here a similar approach and Fig.~\ref{fig:lumlumus} shows the results of our work on a such diagram. Besides disk mass (upper left panel), we also group models by characteristic radius (upper middle) and stellar mass (upper right). Additionally, we provide parameter uncertainty distributions for the diagrams (lower panels). Parameter uncertainty for a chosen point on a diagram is defined as a ratio between the maximal and minimal parameter values of models within 10\% range of luminosities of the chosen point
(10\% chosen as it is a typical absolute flux calibration error for ALMA). Given the ranges of values on our grid, the maximum possible uncertainties are 200, 16 and 4 for disk mass, characteristic radius and stellar mass, respectively.}

{Looking first at disk mass (left panels of {Fig.}~\ref{fig:lumlumus}), we see that while the increase in disk mass leads to an increase in luminosities of both lines, it is significantly smaller than the luminosity variation for a fixed mass (factor of two versus an order of magnitude). The result is an overlap of models of different mass up to models of lowest and highest mass meeting each other. This can be seen in the uncertainty distribution as well, where in the middle of the distribution there exists a spot with uncertainty of at least two orders of magnitude. While the spot does not seem large, it is important to note that its size is also determined by parameter ranges we use. Considering disks with a wider range of mass and size will inevitably expand this spot.}

{Moving to disk characteristic radius (middle panels of {Fig.}~\ref{fig:lumlumus}), models seem to group by this parameter much more distinctively compared to disk mass. This is because the $^{13}$CO line luminosity increases with radius for most of the models while for C$^{18}$O line, the behavior is more mixed. The area occupied by the models is still rather localized, so on the uncertainty distribution we see in the middle a thin strip of maximum possible uncertainty of 16. Substituting $^{13}$CO with CO, luminosity of which increases with radius in all models, gives a better separation of models of different radii with maximum and average uncertainties, that are seen there, 4 and 2, respectively (right panel of Fig.~\ref{fig:lumlumobs}). The overlap increases, however, towards lower disk masses and uncertainty can reach 8 in some isolated points. It is possible that with disk mass range expanded towards lower values the uncertainty will increase there even more as CO will get more optically thin.}

{Lastly, stellar mass (right panels of {Fig.}~\ref{fig:lumlumus}) cannot be meaningfully differentiated in a luminosity diagram. Most of the uncertainty distribution is occupied by a maximum possible value of 4.}

{Concluding this analysis and considering the parameter grid limitations, using a combination of C$^{18}$O and $^{13}$CO line luminosities can allow to determine the disk mass within a factor of 100-50, characteristic radius within a factor of 16-8 and stellar mass within a factor of 4. Using CO instead of $^{13}$CO improves the uncertainty of the characteristic radius estimate to a factor of 4-2.} Adding the variability inherent to T~Tauri-type stars and outbursting events such as EXor and FUor outbursts, the scatter and uncertainty would increase even more.

{However, the estimates can be improved by using additional information and parameter constraints from other wavelengths (stellar mass from the optical, disk mass from the dust continuum) and spatially resolved observations (for radius). At the same time, inclination can be safely ignored (Section~\ref{subsec:incl}). Results of our modeling (including $R_{90}$) are compiled in Appendix~\ref{app:model_data} for the use by the community.} Another option is to fit multiple observations with full physico-chemical models, such as in~\cite{2010A&A...518L.125T, 2019ApJ...883...98Z}.

\section{{Discussion}}
\label{sec:discussion}

\begin{figure*}
    \centering
    \includegraphics[width=0.45\textwidth]{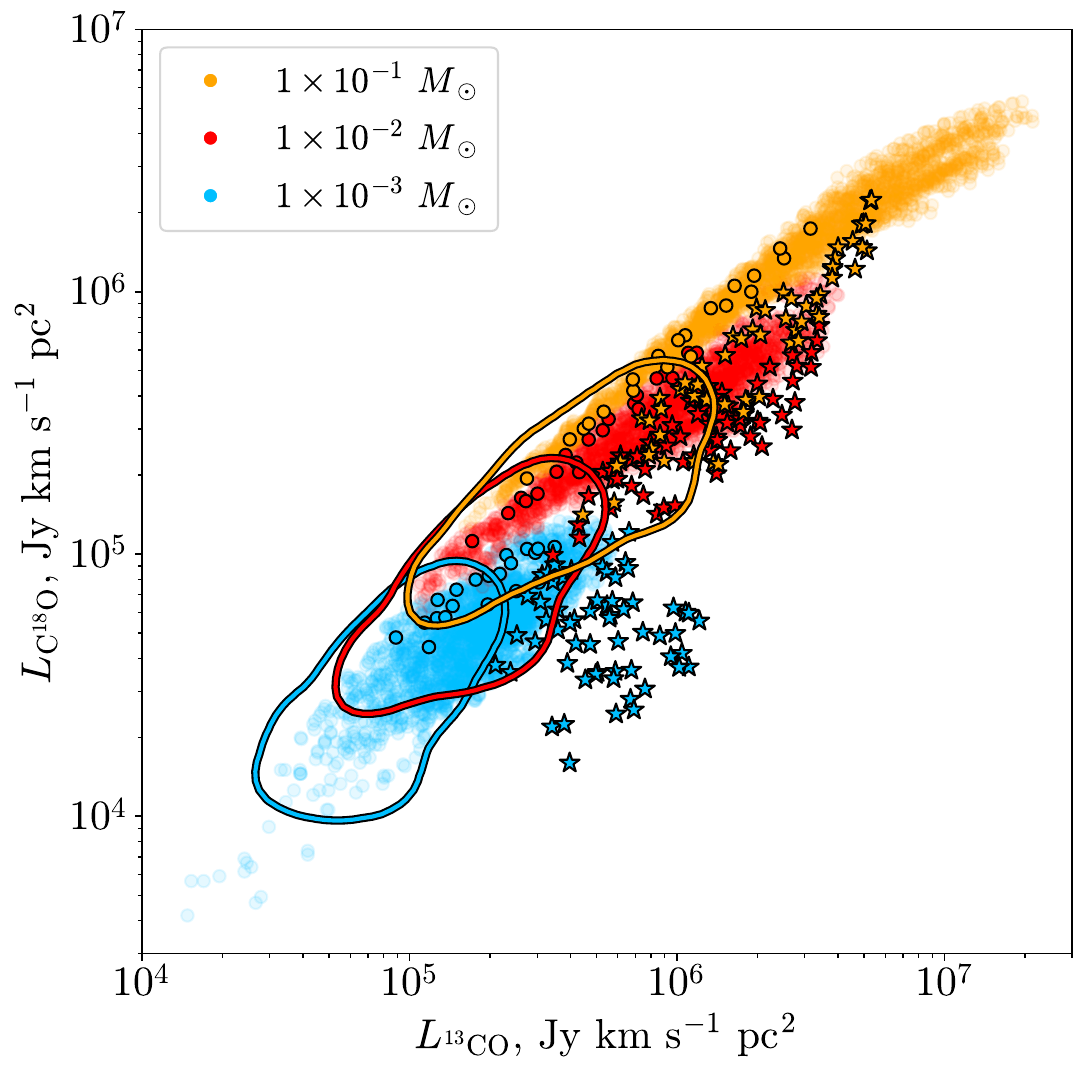}
    \includegraphics[width=0.47\textwidth]{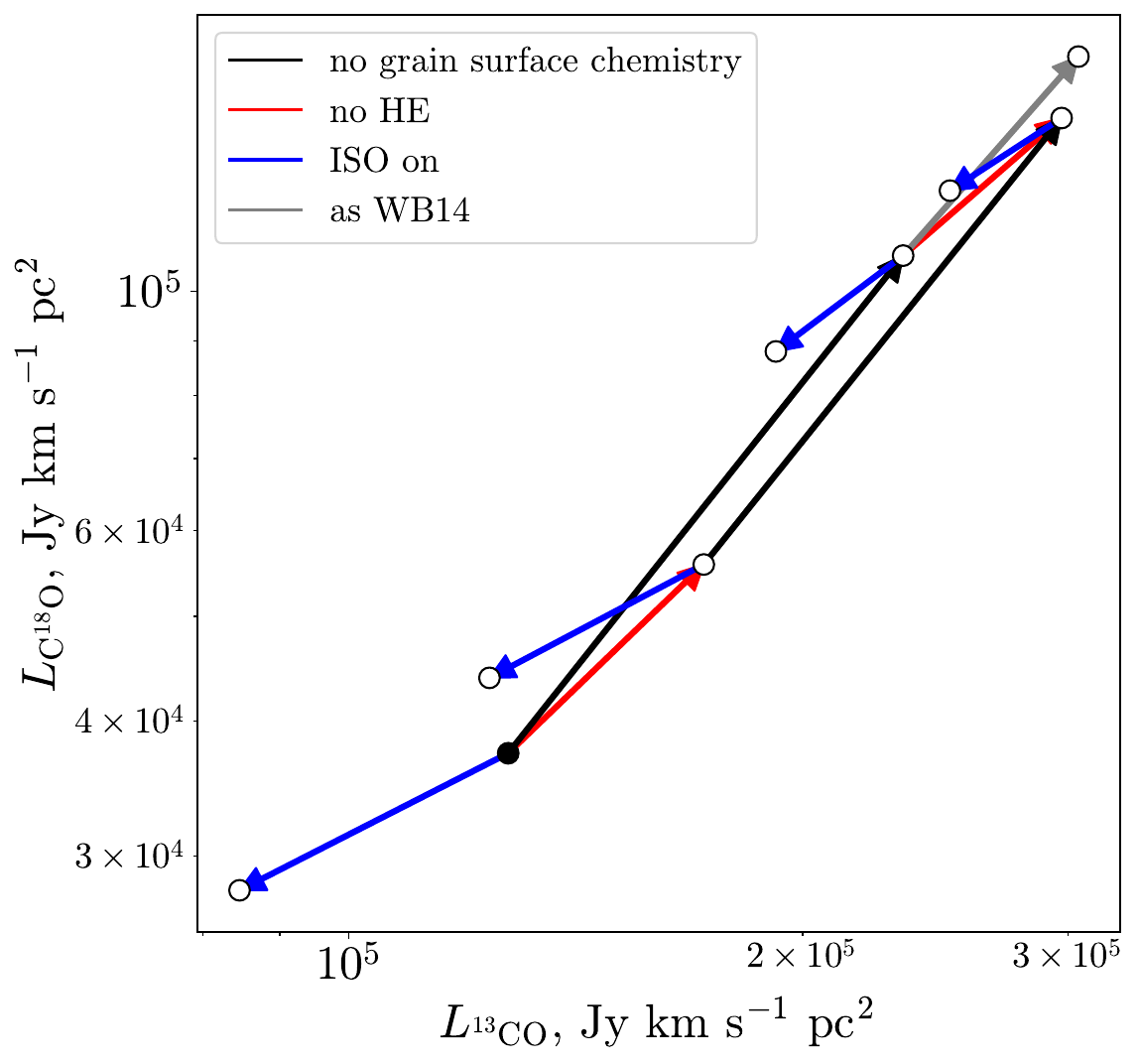}
    \caption{Luminosity diagrams for the C$^{18}$O and $^{13}$CO lines. Left: {translucent} points are results of~\citetalias{2014ApJ...788...59W} assuming CO/C$^{18}$O $=550$ and {traced points} are results of this work using the assumptions from~\citetalias{2014ApJ...788...59W}. {Traced stars are results of~\citetalias{2016A&A...594A..85M} and contours are results of this work with the grain surface chemistry network.} {The points {and contours} are colored according to the disk mass of the model.} Right: {points are values of individual test models. Black point is the {grain surface chemistry} model, white points are modified versions of it. Color of the arrow shows the added modification to the model: grain chemistry turned off (black), parametric vertical structure (no HE, red), isotope-selective photodissociation (ISO on, blue) and \citetalias{2014ApJ...788...59W} assumptions (as WB14, gray). All models have the same parameters: $M_{\rm d} = 0.01\,M_\odot$, $R_{\rm c} = 60$\,au, $M_\star=1.0\,M_\odot$, $L_\star=1.0\,L_\odot$, $T_\star=4000\,$K. Models with parametric vertical structure have $h_{\rm c}=0.1$, $\psi = 0.1$.}}
    \label{fig:lumlumWB14}
\end{figure*}

\begin{figure*}
    \centering
    \includegraphics[width=0.45\textwidth]{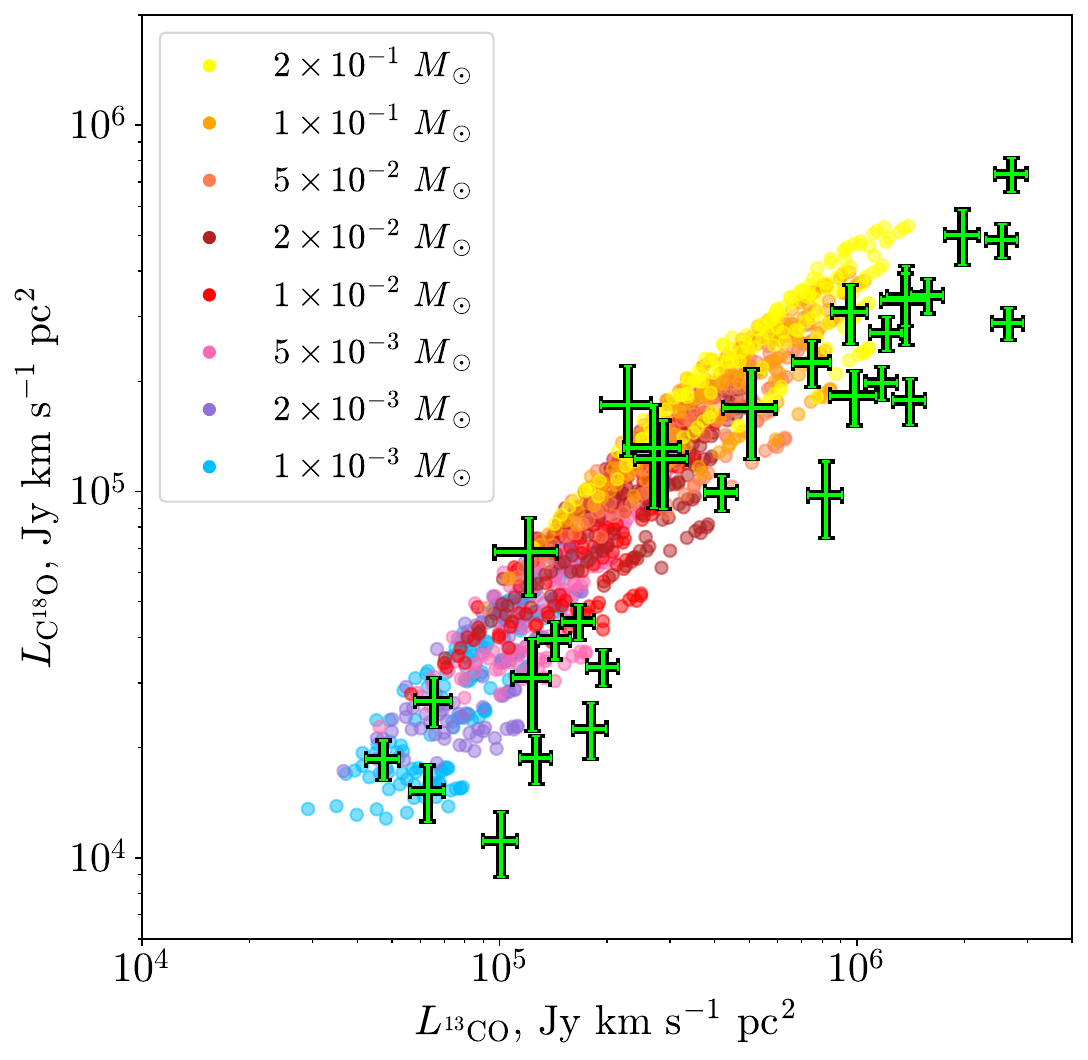}
    \includegraphics[width=0.46\textwidth]{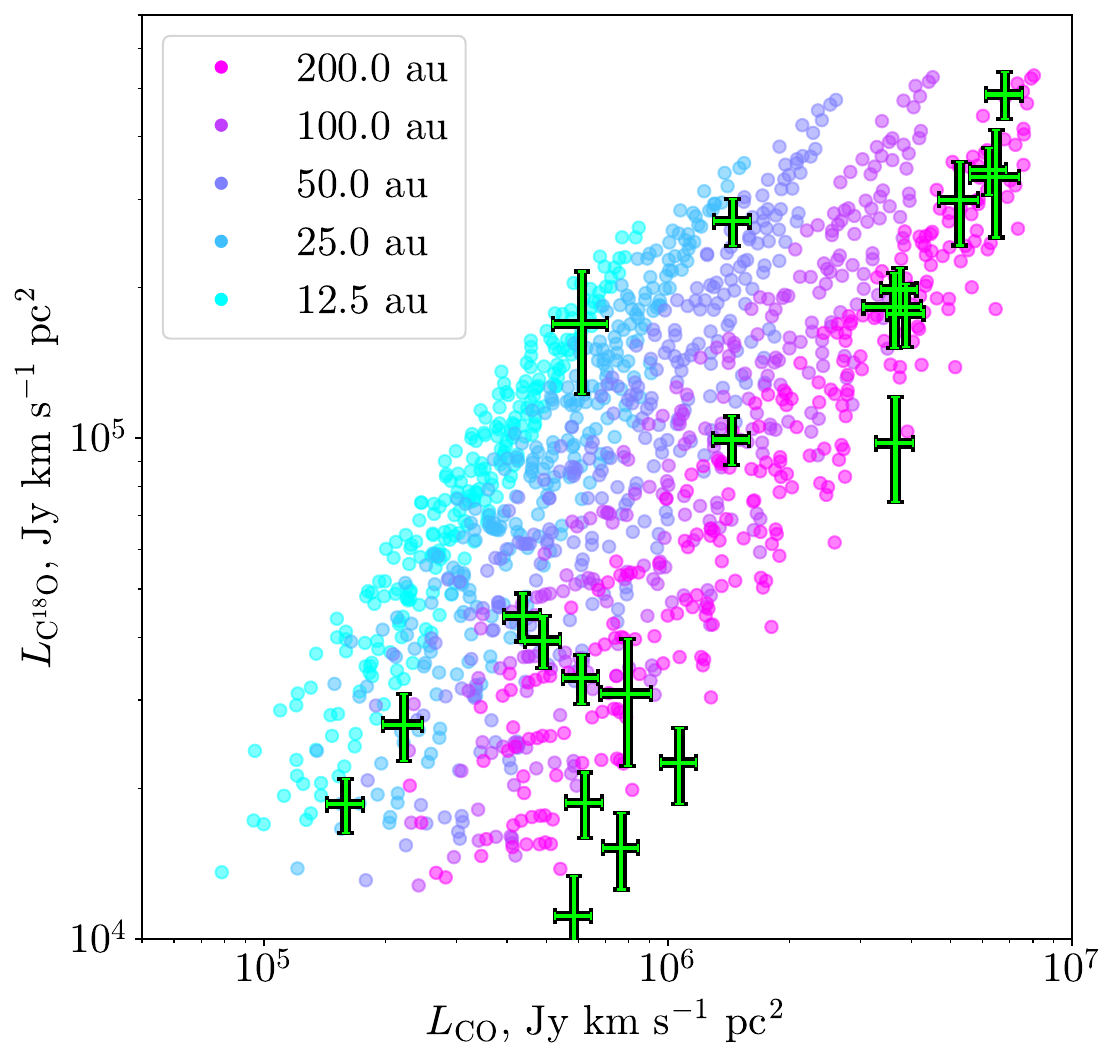}
    \caption{Left: luminosity diagram for the C$^{18}$O and $^{13}$CO lines. {Translucent} points are results of this work with {grain surface chemistry} network {The points are colored according to the disk mass of the model. Markers with errorbars are observations compiled in~\cite{2024MNRAS.527.7652Z} and from~\cite{2016ApJ...828...46A}. Right: luminosity diagram in the C$^{18}$O and CO lines. All points are results of this work with {grain surface chemistry} network and colored according to the $R_{\rm c}$. Markers with errorbars are observations compiled in~\cite{2024MNRAS.527.7652Z}.}}
    \label{fig:lumlumobs}
\end{figure*}

{In this Section we compare our results to previous works {and observations} and analyze how differences between model assumptions cause different outcomes.

\subsection{{Effect of model assumptions}}
\label{sec:discussion_assumptions}

The left panel of Fig.~\ref{fig:lumlumWB14} shows a luminosity diagram with translucent points presenting results of~\citetalias{2014ApJ...788...59W}, {traced points presenting our results with~\citetalias{2014ApJ...788...59W} assumptions,} traced star points presenting results of~\citetalias{2016A&A...594A..85M} and contours {marking the} results of this work with {the grain surface chemistry} network. Compared to~\citetalias{2014ApJ...788...59W} and~\citetalias{2016A&A...594A..85M}, the luminosities of our {grain surface chemistry} models correspond to higher {gas disk} masses, with the shift being about one order of magnitude. This {alleviates} the current problem with CO mass estimation, {that is the masses from CO isotopologues are} by 1-2 orders of magnitude {lower than those derived from other molecular tracers} (e.g., HD)~\citep{2022ApJ...926L...2T}. There is also a higher degree of overlap between our models of different mass which we explore separately in Section~\ref{subsec:linecombo}. 

{At the same time, our results with the assumptions of~\citetalias{2014ApJ...788...59W} align with their results quite well. In Fig.~\ref{fig:mdi60} we also show how flux under~\citetalias{2014ApJ...788...59W} assumptions depends on disk parameters in comparison to the results of our grain surface chemistry network. Comparing the two approaches, emission in the main CO isotopologue line is the least affected. While the dependence on radius for low-mass disks is noticeably different, the difference in flux between two models with the same parameters is within a factor of two. Emission in the $^{13}$CO and C$^{18}$O lines is affected more significantly, and the differences in flux between models with and without chemistry reach a factor of 3 and 5, respectively. Moreover, for all disk masses, the flux dependences on disk radius and stellar mass are completely different.}

{The line fluxes predicted by our models are different from those obtained in previous studies, as we use different model assumptions. The main differences are as follows. First, unlike~\citetalias{2014ApJ...788...59W}, both we and~\citetalias{2016A&A...594A..85M} use chemical network to calculate the CO abundances, which accounts for processes beyond thermal desorption, freeze-out and photo-dissociation. Second, our grain chemistry includes more complex processes than hydrogenation included by~\citetalias{2016A&A...594A..85M}, particularly the CO chemical depletion, through surface conversion to CO$_2$~\citep{2018A&A...618A.182B}. Third, the disk vertical structure in our model (and in~\citetalias{2014ApJ...788...59W}) is determined by hydrostatic equilibrium and calculated self-consistently with temperature, instead of being parametrically set. Finally, our model lacks the inclusion of the isotope-selective processes of~\citetalias{2016A&A...594A..85M}. 
Below we will disentangle the effects of these assumptions on the line fluxes, using different intermediate models that share some of the assumptions.}

{We use a baseline model with the same system parameters: $M_{\rm d} = 0.01\,M_\odot$, $R_{\rm c} = 60$\,au, $M_\star=1.0\,M_\odot$, $L_\star=1.0\,L_\odot$, $T_\star=4000\,$K. One deviation from this model is the vertical structure: we use the same Gaussian parametrization as~\citetalias{2016A&A...594A..85M} and refer to such models as ``no HE'' (no hydrostatic equilibrium) models. Another model variation is the models with {reactions on dust grain surface turned off}. We also explore the~\citetalias{2014ApJ...788...59W} approach with no chemistry. Finally, we test the effect of isotopologue selective photodissociation by adopting different depletion levels of $^{13}$CO and C$^{18}$O based of shielding factors by~\citet{2009A&A...503..323V}. This allows to assess the effect of isotopologue-selective processes in the disk upper layers at post-processing.} 

{Line luminosities of these test models are presented in the {left} panel of Fig.~\ref{fig:lumlumWB14}. The model marked by a black point is our {grain surface chemistry} model. There are two main variations of this model, marked by red and black arrows: one with parametric vertical structure ($h_{\rm c}=0.1$, $\psi = 0.1$) and one with grain surface {reactions} off. Both parametric structure and no grain surface {reactions} give an increase in flux for both lines, with surface chemistry having a stronger effect (a factor of 2-3 increase versus a factor of 1.5). Combining both modifications further increases the flux. These effects are illustrated in more detail in~Appendix~\ref{app:test_models}. {Finally}, there is a model with assumptions from~\citetalias{2014ApJ...788...59W} that has the highest flux among all test models, as it accounts for minimum CO depletion level.} 

{Additionally, each model except for the one with~\citetalias{2014ApJ...788...59W} assumptions was tested for the effect of isotopologue-selective photodissociation (marked by blue arrows). We assume that the abundances of CO isotopologues in the upper disk layers are determined by photodissociation~\citepalias{2016A&A...594A..85M}, and the relative species abundances are inversely proportional to the photodissociation rates. In these models, we artificially reduce the fraction of the isotopologue by a factor of $\theta_{\rm CO}/\theta_{\rm ^{13}CO}$ and $\theta_{\rm CO}/\theta_{\rm C^{18}O}$, in the layer with $\theta_{\rm ^{13}CO}$ or $\theta_{\rm C^{18}O}$ above $10^{-2}$. Here $\theta_{\rm X}$ is the shielding factor for the molecule~X, calculated from the parameterization provided by \citet{2009A&A...503..323V}, see their Table~5. {The values of the applied reduction factor vary across the disk, ranging from~$\approx1$ in the upper layers where shielding is not yet efficient to~$\approx10$ deeper in the molecular layer.} In all cases inclusion of this approximation leads to a decrease in the rare isotopologue abundances in the upper layers {(by a typical factor of $\sim2-3$)}, and consequently to the decrease in their fluxes. The difference is around a factor $1.2-1.5$ for both isotopologues, which is much lower than the effect of chemical processes. However, selective photodissociation is not the only effect of isotopologue chemistry. Other isotopologue selective chemical reactions can also increase the abundances, e.g. by a factor of $\sim2$ for $^{13}$CO, as was shown by~\citet{2014A&A...572A..96M}. However, they are hard to assess without running full isotopologue-selective chemical network, so we limit our analysis to the effect of photodissociation. The fluxes in the~\citetalias{2016A&A...594A..85M} models are also higher than in our simulations (see {Fig.}~\ref{fig:lumlumobs}) due to their use of parametric vertical structure, which increases the fluxes (red arrows in {Fig.}~\ref{fig:lumlumWB14}).}

{Observed fluxes of CO isotopologues can be affected by other physical processes unaccounted for in our modeling. One of the most important factors is the thermal structure of the disk, which can be different from the adopted parametrization. First, the disk thermal structure can have a significant effect on CO freeze-out and the level of CO depletion. For example, the existence of a thick vertically isothermal layer around the disk midplane has been suggested based on the observational data and would also affect the emission of CO and its isotopologues \citep{2019ApJ...882..160Q,2024ApJ...977...60Q}. Vertical mixing, radial drift and dust settling can also enhance CO depletion \citep{2019ApJ...883...98Z,2020ApJ...899..134K}, even without processing to CO$_2$ \citep{2022ApJ...927..206V}. Different adsorption rates to dust grains of different sizes can also increase the CO depletion from the gas phase \citep{2022NatAs...6.1147P}. Similarly, adsorption favors larger grains due to their lower temperatures \citep{2025A&A...694A.213K}, which could also contribute to the level of depletion.}

{Substructures, such as rings or spirals can form in the disk even before 1\,Myr age, affecting its physical structure and consequently the chemical depletion rates. At longer timescales, other processes can become important, such as disk photoevaporation, stellar evolution, disk-planet interaction and most notably dust evolution. Gas-phase CO can also be significantly depleted by magnetic winds (Jonczyk et al., in prep). Each of these effects can affect the abundances of CO and the isotopologues and the timescales of chemical reactions. We consider a 1\,Myr old Class~II disk, where these effects can be overlooked to some degree. Nevertheless, the inclusion of the above processes can significantly affect CO abundances and the level of depletion in a particular disk, increasing the degree of uncertainty in the disk parameter determination.}

{Concluding, the difference in the derived luminosities compared to \citetalias{2014ApJ...788...59W} and \citetalias{2016A&A...594A..85M} can be explained by the combination of chemistry, particularly on dust surface, vertical disk structure and isotopologue-selective processes. We show that the most significant contribution comes from surface chemistry, in particular the mechanism of CO conversion to CO$_2$ on dust.}

\subsection{{Measuring disk masses {and radii}}}
\label{sec:discussion_masses}

{In Fig.~\ref{fig:lumlumobs} we plot some observations of T~Tauri disks compiled in~\cite{2024MNRAS.527.7652Z} \citep{1997A&A...317L..55D,2014ApJ...788...59W,2017AJ....153..240A,2021ApJ...913..123G,2021ApJ...911..150P,2024A&A...685A.126S} and from~\cite{2016ApJ...828...46A} as markers with errorbars. We see that the majority of objects {overlap with our data points} with a few having either higher $^{13}$CO luminosity or higher luminosity in both lines. Considering larger and more massive model disks can help reproduce the values observed in these objects. Such massive disks are not included in our modeling, as the assumptions made for the disk structure would not be appropriate for the self-gravitating disks. Besides, masses of the most massive disks can be determined more directly, from their kinematic structure \citep{2024A&A...688A.136V}. Finally, recent FUor outbursts can increase flux in both of these lines~\citep{2024MNRAS.527.7652Z}. Some of the objects in the plot, namely CI~Tau, [MGM2012]~556 and IM~Lup, are considered to be post-FUor candidates based on the elevated C$^{18}$O/CO and $^{13}$CO/CO flux ratios so their parameters can be overestimated using this method.}

{On the right panel of Fig.~\ref{fig:lumlumobs} we present a similar luminosity diagram but with $^{13}$CO substituted by CO and only with our {grain surface chemistry} results. Markers with errorbars are observations compiled in~\citet{2024MNRAS.527.7652Z}. Here we also see our model set covering most of the observational values making it possible to give some estimates of the characteristic radius. In Appendix~\ref{app:model_data} we present a table with these estimates for both $R_{90}$ and $R_{\rm c}$. It is important to note that, per~\cite{2024MNRAS.527.7652Z} predictions, past outburst can lead to a lower radius estimate using this method.}

{Our results are directly related to the problem of measuring disk gas masses and radii, which are one of the key goals of the upcoming ALMA large programs AGE-PRO~\citep{2024AAS...24410907Z} and DECO~\citep{2025AAS...24520604L}. AGE-PRO and DECO observations are planned to be converted into masses using a method by~\citet{2022ApJ...926L...2T}. This method involves obtaining disk structure from SEDs and resolved observations, then constraining possible values of CO abundance and disk mass using C$^{18}$O flux and finally determining both of them from N$_2$H$^+$/C$^{18}$O flux ratio. Their model uses a chemical network similar to~\citetalias{2016A&A...594A..85M} that does not describe the chemical depletion of CO on the dust surface. Instead, they account for any unspecified source of CO depletion by setting global gas-phase CO abundance as a fitting parameter. However, this may not be sufficient as flux is sensitive to the thermal profile of the emitting area and as we show above and in Appendix~\ref{app:test_models} it depends on the presence of surface chemistry. We would like to emphasize here that careful consideration of CO depletion processes is crucial to accuracy of mass estimation methods, and its inclusion to the models can be beneficial for the accuracy of the mass determination.}

{Another method and a whole tool for determining disk mass is the one proposed by~\citet{2023ApJ...954..165D}. Their model chemistry is based on the work by~\citet{2022ApJ...925...49R}, with a more extensive chemical network including some isotopologue-specific reactions. It is difficult to compare our results due to different approaches to the disk parameter space exploration. Beyond disk mass,~\citet{2022ApJ...925...49R} vary only dust-to-gas mass ratio and outer disk radius. We, on the other hand, consider only one value of dust-to-gas ratio and vary characteristic radius instead of the disk outer rim. However, the dependences of isotopologue flux on disk mass is in agreement with ours and of similar amplitude. The simplified model they construct is checked to be roughly in line with their full chemical model so our results are in approximate agreement with it as well. This may be attributed to the inclusion in the model of hydrostatic equilibrium and CO depletion through conversion to CO$_2$ as we have in our models. \citet{2022ApJ...925...49R} conclude that C$^{18}$O can be a good measure of disk gas mass given that there are good constraints on other model parameters from other sources, which is in agreement with the results of the present study.}

\section{Conclusions}\label{sec:concl}

In this work we explored how the line fluxes of various CO isotopologues are affected by varying disk mass and radius, stellar mass and inclination. Using the astrochemical model ANDES~\citep{2013ApJ...766....8A} and the radiative transfer code RADMC-3D~\citep{2012ascl.soft02015D} we simulated flux in CO, $^{13}$CO, C$^{18}$O $J=2-1$ lines on a grid of models of varying disk mass $M_{\rm d}$ and radius $R_{\rm c}$, inclination $i$ and stellar mass $M_\star$. We explored the dependences of flux on inclination and disk mass with other parameters fixed. We provided physical {and chemical} explanations for the influence of each varied parameter. We approximated dependences on disk mass loglinearly and described how coefficients of approximations change with disk radius and stellar mass. {We tested if the combinations of lines can be used to estimate parameters similar to the methodology of~\citet{2014ApJ...788...59W}. Finally,} we compared our results to~\citet{2014ApJ...788...59W},~\citet{2016A&A...594A..85M} {and observations and explored the main reasons behind differences in our results}. Our findings can be summarized as follows:
\begin{enumerate}
    \item Flux in all lines is independent of inclination for $i<60$\textdegree. For $i>60$\textdegree, it decreases with higher inclinations but within a factor of $<2$.
    \item Flux in all lines increases with disk mass. 
    \item CO flux increases with disk radius while $^{13}$CO and C$^{18}$O flux dependence on it is more complex and depends both on disk and stellar mass due to a combined effect of CO depletion area, temperature and photodissociation on these more optically thin lines.
    \item Due to these dependences, {use of just one line cannot provide reliable estimates of any parameter. However, having prior constraints on some parameters can improve the quality of the estimate}.
    \item Using {$^{13}$CO-C$^{18}$O} line combination as per~\citet{2014ApJ...788...59W} cannot produce reliable estimates of disk mass {or radius} as our grid of models disks of different mass { and radius heavily} overlap in luminosity. {Using CO-C$^{18}$O line combination can give an estimate of disk characteristic radius within a factor of 2-4.}
    \item {Chemical depletion of CO leads to CO absent in both gas and ice phases at temperatures below $34-65$\,K and above the freeze-out temperature of $17-29$\,K, with the certain values depending on the model parameters. It is caused by the transformation to CO$_2$ on the dust grain surface.}
    \item The inclusion of surface chemistry {significantly reduces $^{13}$CO and C$^{18}$O isotopologue flux as their emission traces the region of chemical CO depletion. This reduction in flux is a major source of a systematic underestimation of disk mass in previous works. A choice in disk vertical structure description also impacts the flux. A structure from hydrostatic equilibrium gives a lower $^{13}$CO and C$^{18}$O flux compared to parametric prescription.}
\end{enumerate}

\section*{Data availability}

The model parameters, $R_{90}$ and flux data shown in Table~\ref{tab:modeldata} is available at \url{https://doi.org/10.5281/zenodo.15690896}. Additional data can be provided
upon request to the corresponding authors.

\begin{acknowledgements}
      {We are thankful to the anonymous referee for their constructive comments and suggestions that helped to improve the manuscript.} L. Zwicky, P. \'Abrah\'am, and \'A. K\'osp\'al acknowledge financial support from the Hungarian NKFIH OTKA project no. K-147380. T. Molyarova was supported by the Royal Society, award numbers URF\textbackslash R1\textbackslash 211799 and RF\textbackslash ERE\textbackslash 231082. The research leading to these results has received funding from the European Union’s Horizon 2020 research and innovation programme under Grant Agreement 101004719 (ORP). {This work was also supported by the NKFIH NKKP grant ADVANCED 149943 and the NKFIH excellence grant TKP2021-NKTA-64. Project no.149943 has been implemented with the support provided by the Ministry of Culture and Innovation of Hungary from the National Research, Development and Innovation Fund, financed under the NKKP ADVANCED funding scheme. This publication is based upon work from COST Action PLANETS CA22133, supported by COST (European Cooperation in Science and Technology).}
    
\end{acknowledgements}

\bibliographystyle{aa} 
\bibliography{References}

\begin{thebibliography}{60}
\expandafter\ifx\csname natexlab\endcsname\relax\def\natexlab#1{#1}\fi

\bibitem[{{Akimkin} {et~al.}(2013){Akimkin}, {Zhukovska}, {Wiebe}, {Semenov}, {Pavlyuchenkov}, {Vasyunin}, {Birnstiel}, \& {Henning}}]{2013ApJ...766....8A}
{Akimkin}, V., {Zhukovska}, S., {Wiebe}, D., {et~al.} 2013, \apj, 766, 8

\bibitem[{{Andrews} \& {Williams}(2005)}]{2005ApJ...631.1134A}
{Andrews}, S.~M. \& {Williams}, J.~P. 2005, \apj, 631, 1134

\bibitem[{{Ansdell} {et~al.}(2017){Ansdell}, {Williams}, {Manara}, {Miotello}, {Facchini}, {van der Marel}, {Testi}, \& {van Dishoeck}}]{2017AJ....153..240A}
{Ansdell}, M., {Williams}, J.~P., {Manara}, C.~F., {et~al.} 2017, \aj, 153, 240

\bibitem[{{Ansdell} {et~al.}(2016){Ansdell}, {Williams}, {van der Marel}, {Carpenter}, {Guidi}, {Hogerheijde}, {Mathews}, {Manara}, {Miotello}, {Natta}, {Oliveira}, {Tazzari}, {Testi}, {van Dishoeck}, \& {van Terwisga}}]{2016ApJ...828...46A}
{Ansdell}, M., {Williams}, J.~P., {van der Marel}, N., {et~al.} 2016, \apj, 828, 46

\bibitem[{{Bergin} {et~al.}(2013){Bergin}, {Cleeves}, {Gorti}, {Zhang}, {Blake}, {Green}, {Andrews}, {Evans}, {Henning}, {{\"O}berg}, {Pontoppidan}, {Qi}, {Salyk}, \& {van Dishoeck}}]{2013Natur.493..644B}
{Bergin}, E.~A., {Cleeves}, L.~I., {Gorti}, U., {et~al.} 2013, \nat, 493, 644

\bibitem[{{Bitsch} {et~al.}(2015){Bitsch}, {Lambrechts}, \& {Johansen}}]{2015A&A...582A.112B}
{Bitsch}, B., {Lambrechts}, M., \& {Johansen}, A. 2015, \aap, 582, A112

\bibitem[{{Bohlin} {et~al.}(1978){Bohlin}, {Savage}, \& {Drake}}]{1978ApJ...224..132B}
{Bohlin}, R.~C., {Savage}, B.~D., \& {Drake}, J.~F. 1978, \apj, 224, 132

\bibitem[{{Booth} {et~al.}(2021){Booth}, {Walsh}, {Terwisscha van Scheltinga}, {van Dishoeck}, {Ilee}, {Hogerheijde}, {Kama}, \& {Nomura}}]{2021NatAs...5..684B}
{Booth}, A.~S., {Walsh}, C., {Terwisscha van Scheltinga}, J., {et~al.} 2021, Nature Astronomy, 5, 684

\bibitem[{{Bosman} {et~al.}(2018){Bosman}, {Walsh}, \& {van Dishoeck}}]{2018A&A...618A.182B}
{Bosman}, A.~D., {Walsh}, C., \& {van Dishoeck}, E.~F. 2018, \aap, 618, A182

\bibitem[{{Bruderer} {et~al.}(2009){Bruderer}, {Doty}, \& {Benz}}]{2009ApJS..183..179B}
{Bruderer}, S., {Doty}, S.~D., \& {Benz}, A.~O. 2009, \apjs, 183, 179

\bibitem[{{Cuppen} {et~al.}(2017){Cuppen}, {Walsh}, {Lamberts}, {Semenov}, {Garrod}, {Penteado}, \& {Ioppolo}}]{2017SSRv..212....1C}
{Cuppen}, H.~M., {Walsh}, C., {Lamberts}, T., {et~al.} 2017, \ssr, 212, 1

\bibitem[{{Deng} {et~al.}(2023){Deng}, {Ruaud}, {Gorti}, \& {Pascucci}}]{2023ApJ...954..165D}
{Deng}, D., {Ruaud}, M., {Gorti}, U., \& {Pascucci}, I. 2023, \apj, 954, 165

\bibitem[{{Dullemond} {et~al.}(2012){Dullemond}, {Juhasz}, {Pohl}, {Sereshti}, {Shetty}, {Peters}, {Commercon}, \& {Flock}}]{2012ascl.soft02015D}
{Dullemond}, C.~P., {Juhasz}, A., {Pohl}, A., {et~al.} 2012, {RADMC-3D: A multi-purpose radiative transfer tool}, Astrophysics Source Code Library, record ascl:1202.015

\bibitem[{{Dutrey} {et~al.}(1997){Dutrey}, {Guilloteau}, \& {Guelin}}]{1997A&A...317L..55D}
{Dutrey}, A., {Guilloteau}, S., \& {Guelin}, M. 1997, \aap, 317, L55

\bibitem[{{Eistrup} {et~al.}(2016){Eistrup}, {Walsh}, \& {van Dishoeck}}]{2016A&A...595A..83E}
{Eistrup}, C., {Walsh}, C., \& {van Dishoeck}, E.~F. 2016, \aap, 595, A83

\bibitem[{{Evans} {et~al.}(2025){Evans}, {Booth}, {Walsh}, {Ilee}, {Keyte}, {Law}, {Leemker}, {Notsu}, {{\"O}berg}, {Temmink}, \& {van der Marel}}]{2025arXiv250204957E}
{Evans}, L., {Booth}, A.~S., {Walsh}, C., {et~al.} 2025, arXiv e-prints, arXiv:2502.04957

\bibitem[{{Favre} {et~al.}(2013){Favre}, {Cleeves}, {Bergin}, {Qi}, \& {Blake}}]{2013ApJ...776L..38F}
{Favre}, C., {Cleeves}, L.~I., {Bergin}, E.~A., {Qi}, C., \& {Blake}, G.~A. 2013, \apjl, 776, L38

\bibitem[{{Grant} {et~al.}(2021){Grant}, {Espaillat}, {Wendeborn}, {Tobin}, {Mac{\'\i}as}, {Rilinger}, {Ribas}, {Megeath}, {Fischer}, {Calvet}, \& {Hee Kim}}]{2021ApJ...913..123G}
{Grant}, S.~L., {Espaillat}, C.~C., {Wendeborn}, J., {et~al.} 2021, \apj, 913, 123

\bibitem[{{Harsono} {et~al.}(2015){Harsono}, {Bruderer}, \& {van Dishoeck}}]{2015A&A...582A..41H}
{Harsono}, D., {Bruderer}, S., \& {van Dishoeck}, E.~F. 2015, \aap, 582, A41

\bibitem[{{Kalv{\={a}}ns}(2025)}]{2025A&A...694A.213K}
{Kalv{\={a}}ns}, J. 2025, \aap, 694, A213

\bibitem[{{Kama} {et~al.}(2020){Kama}, {Trapman}, {Fedele}, {Bruderer}, {Hogerheijde}, {Miotello}, {van Dishoeck}, {Clarke}, \& {Bergin}}]{2020A&A...634A..88K}
{Kama}, M., {Trapman}, L., {Fedele}, D., {et~al.} 2020, \aap, 634, A88

\bibitem[{{Krijt} {et~al.}(2020){Krijt}, {Bosman}, {Zhang}, {Schwarz}, {Ciesla}, \& {Bergin}}]{2020ApJ...899..134K}
{Krijt}, S., {Bosman}, A.~D., {Zhang}, K., {et~al.} 2020, \apj, 899, 134

\bibitem[{{Law} {et~al.}(2025){Law}, {Cleeves}, \& {DECO Team}}]{2025AAS...24520604L}
{Law}, C., {Cleeves}, L.~I., \& {DECO Team}. 2025, in American Astronomical Society Meeting Abstracts, Vol. 245, American Astronomical Society Meeting Abstracts, 206.04

\bibitem[{{Leemker} {et~al.}(2021){Leemker}, {van't Hoff}, {Trapman}, {van Gelder}, {Hogerheijde}, {Ru{\'\i}z-Rodr{\'\i}guez}, \& {van Dishoeck}}]{2021A&A...646A...3L}
{Leemker}, M., {van't Hoff}, M.~L.~R., {Trapman}, L., {et~al.} 2021, \aap, 646, A3

\bibitem[{{Mathis} {et~al.}(1977){Mathis}, {Rumpl}, \& {Nordsieck}}]{1977ApJ...217..425M}
{Mathis}, J.~S., {Rumpl}, W., \& {Nordsieck}, K.~H. 1977, \apj, 217, 425

\bibitem[{{McClure} {et~al.}(2016){McClure}, {Bergin}, {Cleeves}, {van Dishoeck}, {Blake}, {Evans}, {Green}, {Henning}, {{\"O}berg}, {Pontoppidan}, \& {Salyk}}]{2016ApJ...831..167M}
{McClure}, M.~K., {Bergin}, E.~A., {Cleeves}, L.~I., {et~al.} 2016, \apj, 831, 167

\bibitem[{{Miotello} {et~al.}(2014){Miotello}, {Bruderer}, \& {van Dishoeck}}]{2014A&A...572A..96M}
{Miotello}, A., {Bruderer}, S., \& {van Dishoeck}, E.~F. 2014, \aap, 572, A96

\bibitem[{{Miotello} {et~al.}(2023){Miotello}, {Kamp}, {Birnstiel}, {Cleeves}, \& {Kataoka}}]{2023ASPC..534..501M}
{Miotello}, A., {Kamp}, I., {Birnstiel}, T., {Cleeves}, L.~C., \& {Kataoka}, A. 2023, in Astronomical Society of the Pacific Conference Series, Vol. 534, Protostars and Planets VII, ed. S.~{Inutsuka}, Y.~{Aikawa}, T.~{Muto}, K.~{Tomida}, \& M.~{Tamura}, 501

\bibitem[{{Miotello} {et~al.}(2016){Miotello}, {van Dishoeck}, {Kama}, \& {Bruderer}}]{2016A&A...594A..85M}
{Miotello}, A., {van Dishoeck}, E.~F., {Kama}, M., \& {Bruderer}, S. 2016, \aap, 594, A85

\bibitem[{{Molyarova} {et~al.}(2018){Molyarova}, {Akimkin}, {Semenov}, {{\'A}brah{\'a}m}, {Henning}, {K{\'o}sp{\'a}l}, {Vorobyov}, \& {Wiebe}}]{2018ApJ...866...46M}
{Molyarova}, T., {Akimkin}, V., {Semenov}, D., {et~al.} 2018, \apj, 866, 46

\bibitem[{{Molyarova} {et~al.}(2017){Molyarova}, {Akimkin}, {Semenov}, {Henning}, {Vasyunin}, \& {Wiebe}}]{2017ApJ...849..130M}
{Molyarova}, T., {Akimkin}, V., {Semenov}, D., {et~al.} 2017, \apj, 849, 130

\bibitem[{{Mordasini} {et~al.}(2012){Mordasini}, {Alibert}, {Benz}, {Klahr}, \& {Henning}}]{2012A&A...541A..97M}
{Mordasini}, C., {Alibert}, Y., {Benz}, W., {Klahr}, H., \& {Henning}, T. 2012, \aap, 541, A97

\bibitem[{{Nomura} {et~al.}(2016){Nomura}, {Tsukagoshi}, {Kawabe}, {Ishimoto}, {Okuzumi}, {Muto}, {Kanagawa}, {Ida}, {Walsh}, {Millar}, \& {Bai}}]{2016ApJ...819L...7N}
{Nomura}, H., {Tsukagoshi}, T., {Kawabe}, R., {et~al.} 2016, \apjl, 819, L7

\bibitem[{{{\"O}berg} {et~al.}(2023){{\"O}berg}, {Facchini}, \& {Anderson}}]{2023ARA&A..61..287O}
{{\"O}berg}, K.~I., {Facchini}, S., \& {Anderson}, D.~E. 2023, \araa, 61, 287

\bibitem[{{{\"O}berg} {et~al.}(2021){{\"O}berg}, {Guzm{\'a}n}, {Walsh}, {Aikawa}, {Bergin}, {Law}, {Loomis}, {Alarc{\'o}n}, {Andrews}, {Bae}, {Bergner}, {Boehler}, {Booth}, {Bosman}, {Calahan}, {Cataldi}, {Cleeves}, {Czekala}, {Furuya}, {Huang}, {Ilee}, {Kurtovic}, {Le Gal}, {Liu}, {Long}, {M{\'e}nard}, {Nomura}, {P{\'e}rez}, {Qi}, {Schwarz}, {Sierra}, {Teague}, {Tsukagoshi}, {Yamato}, {van't Hoff}, {Waggoner}, {Wilner}, \& {Zhang}}]{2021ApJS..257....1O}
{{\"O}berg}, K.~I., {Guzm{\'a}n}, V.~V., {Walsh}, C., {et~al.} 2021, \apjs, 257, 1

\bibitem[{{Padovani} {et~al.}(2018){Padovani}, {Ivlev}, {Galli}, \& {Caselli}}]{2018A&A...614A.111P}
{Padovani}, M., {Ivlev}, A.~V., {Galli}, D., \& {Caselli}, P. 2018, \aap, 614, A111

\bibitem[{{Pegues} {et~al.}(2021){Pegues}, {{\"O}berg}, {Bergner}, {Huang}, {Pascucci}, {Teague}, {Andrews}, {Bergin}, {Cleeves}, {Guzm{\'a}n}, {Long}, {Qi}, \& {Wilner}}]{2021ApJ...911..150P}
{Pegues}, J., {{\"O}berg}, K.~I., {Bergner}, J.~B., {et~al.} 2021, \apj, 911, 150

\bibitem[{{Powell} {et~al.}(2022){Powell}, {Gao}, {Murray-Clay}, \& {Zhang}}]{2022NatAs...6.1147P}
{Powell}, D., {Gao}, P., {Murray-Clay}, R., \& {Zhang}, X. 2022, Nature Astronomy, 6, 1147

\bibitem[{{Qi} {et~al.}(2019){Qi}, {{\"O}berg}, {Espaillat}, {Robinson}, {Andrews}, {Wilner}, {Blake}, {Bergin}, \& {Cleeves}}]{2019ApJ...882..160Q}
{Qi}, C., {{\"O}berg}, K.~I., {Espaillat}, C.~C., {et~al.} 2019, \apj, 882, 160

\bibitem[{{Qi} \& {Wilner}(2024)}]{2024ApJ...977...60Q}
{Qi}, C. \& {Wilner}, D.~J. 2024, \apj, 977, 60

\bibitem[{{Ruaud} {et~al.}(2022){Ruaud}, {Gorti}, \& {Hollenbach}}]{2022ApJ...925...49R}
{Ruaud}, M., {Gorti}, U., \& {Hollenbach}, D.~J. 2022, \apj, 925, 49

\bibitem[{{Sch{\"o}ier} {et~al.}(2005){Sch{\"o}ier}, {van der Tak}, {van Dishoeck}, \& {Black}}]{2005A&A...432..369S}
{Sch{\"o}ier}, F.~L., {van der Tak}, F.~F.~S., {van Dishoeck}, E.~F., \& {Black}, J.~H. 2005, \aap, 432, 369

\bibitem[{{Semenov} {et~al.}(2024){Semenov}, {Henning}, {Guilloteau}, {Smirnov-Pinchukov}, {Dutrey}, {Chapillon}, {Pi{\'e}tu}, {Franceschi}, {Schwarz}, {van Terwisga}, {Bouscasse}, {Caselli}, {Ceccarelli}, {Cunningham}, {Fuente}, {Gieser}, {Hsieh}, {Lopez-Sepulcre}, {Segura-Cox}, {Pineda}, {Maureira}, {M{\"o}ller}, {Tafalla}, \& {Valdivia-Mena}}]{2024A&A...685A.126S}
{Semenov}, D., {Henning}, T., {Guilloteau}, S., {et~al.} 2024, \aap, 685, A126

\bibitem[{{Semenov} \& {Wiebe}(2011)}]{2011ApJS..196...25S}
{Semenov}, D. \& {Wiebe}, D. 2011, \apjs, 196, 25

\bibitem[{{Tazzari} {et~al.}(2017){Tazzari}, {Testi}, {Natta}, {Ansdell}, {Carpenter}, {Guidi}, {Hogerheijde}, {Manara}, {Miotello}, {van der Marel}, {van Dishoeck}, \& {Williams}}]{2017A&A...606A..88T}
{Tazzari}, M., {Testi}, L., {Natta}, A., {et~al.} 2017, \aap, 606, A88

\bibitem[{{Testi} {et~al.}(2014){Testi}, {Birnstiel}, {Ricci}, {Andrews}, {Blum}, {Carpenter}, {Dominik}, {Isella}, {Natta}, {Williams}, \& {Wilner}}]{2014prpl.conf..339T}
{Testi}, L., {Birnstiel}, T., {Ricci}, L., {et~al.} 2014, in Protostars and Planets VI, ed. H.~{Beuther}, R.~S. {Klessen}, C.~P. {Dullemond}, \& T.~{Henning}, 339--361

\bibitem[{{Thi} {et~al.}(2010){Thi}, {Mathews}, {M{\'e}nard}, {Woitke}, {Meeus}, {Riviere-Marichalar}, {Pinte}, {Howard}, {Roberge}, {Sandell}, {Pascucci}, {Riaz}, {Grady}, {Dent}, {Kamp}, {Duch{\^e}ne}, {Augereau}, {Pantin}, {Vandenbussche}, {Tilling}, {Williams}, {Eiroa}, {Barrado}, {Alacid}, {Andrews}, {Ardila}, {Aresu}, {Brittain}, {Ciardi}, {Danchi}, {Fedele}, {de Gregorio-Monsalvo}, {Heras}, {Huelamo}, {Krivov}, {Lebreton}, {Liseau}, {Martin-Zaidi}, {Mendigut{\'\i}a}, {Montesinos}, {Mora}, {Morales-Calderon}, {Nomura}, {Phillips}, {Podio}, {Poelman}, {Ramsay}, {Rice}, {Solano}, {Walker}, {White}, \& {Wright}}]{2010A&A...518L.125T}
{Thi}, W.~F., {Mathews}, G., {M{\'e}nard}, F., {et~al.} 2010, \aap, 518, L125

\bibitem[{{Trapman} {et~al.}(2017){Trapman}, {Miotello}, {Kama}, {van Dishoeck}, \& {Bruderer}}]{2017A&A...605A..69T}
{Trapman}, L., {Miotello}, A., {Kama}, M., {van Dishoeck}, E.~F., \& {Bruderer}, S. 2017, \aap, 605, A69

\bibitem[{{Trapman} {et~al.}(2022){Trapman}, {Zhang}, {van't Hoff}, {Hogerheijde}, \& {Bergin}}]{2022ApJ...926L...2T}
{Trapman}, L., {Zhang}, K., {van't Hoff}, M. L.~R., {Hogerheijde}, M.~R., \& {Bergin}, E.~A. 2022, \apjl, 926, L2

\bibitem[{{Van Clepper} {et~al.}(2022){Van Clepper}, {Bergner}, {Bosman}, {Bergin}, \& {Ciesla}}]{2022ApJ...927..206V}
{Van Clepper}, E., {Bergner}, J.~B., {Bosman}, A.~D., {Bergin}, E., \& {Ciesla}, F.~J. 2022, \apj, 927, 206

\bibitem[{{Veronesi} {et~al.}(2024){Veronesi}, {Longarini}, {Lodato}, {Laibe}, {Hall}, {Facchini}, \& {Testi}}]{2024A&A...688A.136V}
{Veronesi}, B., {Longarini}, C., {Lodato}, G., {et~al.} 2024, \aap, 688, A136

\bibitem[{{Visser} {et~al.}(2009){Visser}, {van Dishoeck}, \& {Black}}]{2009A&A...503..323V}
{Visser}, R., {van Dishoeck}, E.~F., \& {Black}, J.~H. 2009, \aap, 503, 323

\bibitem[{{Wakelam} {et~al.}(2015){Wakelam}, {Loison}, {Herbst}, {Pavone}, {Bergeat}, {B{\'e}roff}, {Chabot}, {Faure}, {Galli}, {Geppert}, {Gerlich}, {Gratier}, {Harada}, {Hickson}, {Honvault}, {Klippenstein}, {Le Picard}, {Nyman}, {Ruaud}, {Schlemmer}, {Sims}, {Talbi}, {Tennyson}, \& {Wester}}]{2015ApJS..217...20W}
{Wakelam}, V., {Loison}, J.~C., {Herbst}, E., {et~al.} 2015, \apjs, 217, 20

\bibitem[{{Weidenschilling}(1977)}]{1977Ap&SS..51..153W}
{Weidenschilling}, S.~J. 1977, \apss, 51, 153

\bibitem[{{Williams} \& {Best}(2014)}]{2014ApJ...788...59W}
{Williams}, J.~P. \& {Best}, W. M.~J. 2014, \apj, 788, 59

\bibitem[{{Williams} \& {Cieza}(2011)}]{2011ARA&A..49...67W}
{Williams}, J.~P. \& {Cieza}, L.~A. 2011, \araa, 49, 67

\bibitem[{{Yorke} \& {Bodenheimer}(2008)}]{2008ASPC..387..189Y}
{Yorke}, H.~W. \& {Bodenheimer}, P. 2008, in Astronomical Society of the Pacific Conference Series, Vol. 387, Massive Star Formation: Observations Confront Theory, ed. H.~{Beuther}, H.~{Linz}, \& T.~{Henning}, 189

\bibitem[{{Zhang} {et~al.}(2019){Zhang}, {Bergin}, {Schwarz}, {Krijt}, \& {Ciesla}}]{2019ApJ...883...98Z}
{Zhang}, K., {Bergin}, E.~A., {Schwarz}, K., {Krijt}, S., \& {Ciesla}, F. 2019, \apj, 883, 98

\bibitem[{{Zhang} {et~al.}(2024){Zhang}, {Trapman}, {Pascucci}, {Pinilla}, {Perez}, {Cieza}, \& {Carpenter}}]{2024AAS...24410907Z}
{Zhang}, K., {Trapman}, L., {Pascucci}, I., {et~al.} 2024, in American Astronomical Society Meeting Abstracts, Vol. 244, American Astronomical Society Meeting Abstracts, 109.07

\bibitem[{{Zwicky} {et~al.}(2024){Zwicky}, {Molyarova}, {Akimkin}, {Smirnov-Pinchukov}, {Semenov}, {K{\'o}sp{\'a}l}, \& {{\'A}brah{\'a}m}}]{2024MNRAS.527.7652Z}
{Zwicky}, L., {Molyarova}, T., {Akimkin}, V., {et~al.} 2024, \mnras, 527, 7652

\end{thebibliography}

\begin{appendix}
\onecolumn
\section{{Model data and estimates}}\label{app:model_data}
{Table~\ref{tab:modeldata} contains all the computed {grain surface chemistry} models with their parameters, fluxes and $R_{90}$. Table~\ref{tab:obs_table} presents our estimates of $R_{90}$ and $R_{\rm c}$ for a subset of observations compiled in~\cite{2024MNRAS.527.7652Z}. Estimates are obtained by taking minimum and maximum parameter values of the models within errorbars of the objects. If there are no models within object errorbars, an estimate cannot be made and such objects are not listed in the table.}

\begin{table*}[h!]
    \centering
    \caption{{Parameters, $R_{90}$ and fluxes of the grid of models.}}
    \begin{adjustbox}{width=0.8\textwidth,center}
    \begin{tabular}{ccccccccc}
\hline \hline
Model & $M_{\rm d}$ & $R_{\rm c}$ & $R_{90}$ & $M_\star$ & $i$ & $F_{\rm CO}$ & $F_{\rm ^{13}CO}$ & $F_{\rm C^{18}O}$ \\
 & $M_\odot $ & au & au & $M_\odot$ & deg & mJy km/s & mJy km/s & mJy km/s \\ \hline
1 & 0.001 & 12.5 & 39.0 & 0.5 & 0 & 450.64 & 139.10 & 61.17 \\
 & 0.001 & 12.5 & 39.0 & 0.5 & 30 & 462.46 & 146.19 & 63.01 \\
 ... & ... & ... & ... & ... & ... & ... & ... & ... \\
 160 & 0.200 & 200.0 & 970.0 & 2.0 & 90 & 17901.59 & 3256.77 & 1259.85 \\ \hline
 \multicolumn{9}{l}{\tiny (This table is available in its entirety in machine-readable form.)}
    \end{tabular}
    \label{tab:modeldata}
    \end{adjustbox}
    \tablefoot{
Models of the same parameters except inclination have the same number (first column).
}
\end{table*}

\begin{table}[h]
\caption{{Estimates of $R_{90}$ and $R_{\rm c}$ for a set of observations compiled in~\cite{2024MNRAS.527.7652Z}.}}\label{tab:obs_table}
\begin{adjustbox}{width=0.7\textwidth,center}
\begin{tabular}{cccc}
\hline \hline
Disk & $R_{90}$ & $R_{\rm c}$ & Ref.\\
 & au & au & \\\hline
IM Lup & $\sim970$ & $\sim200$ & \cite{2021ApJS..257....1O}\\
GM Aur* & [831, 990] & $\sim200$ & \\
AS 209 & [277, 501] & [50, 200]  & \\\hline
DM Tau* & [613, 825] & [100, 200] & {\cite{1997A&A...317L..55D}} \\
GG Tau & [547, 924] & [100, 200] & \\\hline
Haro 6-13 & $\sim719$ & $\sim200$ & \cite{2014ApJ...788...59W} \\
TW Hya* & [231, 396] & [100, 200] &  \\\hline
{[MGM2012]} 378 & [831, 990] & $\sim200$ & \cite{2021ApJ...913..123G} \\
{[MGM2012]} 556* & [65, 178] & [12.5, 25] &  \\
{[MGM2012]} 561 & [831, 990] & $\sim200$ &  \\\hline
FP Tau & [72, 85] & $\sim25$ & \cite{2021ApJ...911..150P} \\
J0432+1827 & [118, 224] & [50, 100] &  \\
J1100-7619 & [184, 290] & [50, 200] &  \\
J1545-3417 & [52, 125] & [12.5, 100] &  \\\hline
CI Tau & [132, 277] & [25, 50] & {\cite{2024A&A...685A.126S}} \\
CY Tau & [112, 165] & [25, 100] & \\
DM Tau* & [613, 864] & [100, 200] & \\
DN Tau & $\sim217$ & $\sim100$ & \\\hline
\end{tabular}
\end{adjustbox}
\tablefoot{Sign $\sim$ means there was only a single value of the parameter within the observation errorbars. Maximum possible values of the estimates are 990~and 200\,au for $R_{90}$ and $R_{\rm c}$, respectively. References column indicates the work that the fluxes of objects were obtained from. Disks with an asterisk are transitional disks, the rest are Class~II YSOs.}
\end{table}

\section{{Disentangling stellar mass effects}}\label{app:stellar_effects}
\begin{figure*}
    \centering
    \includegraphics[width=0.33\textwidth]{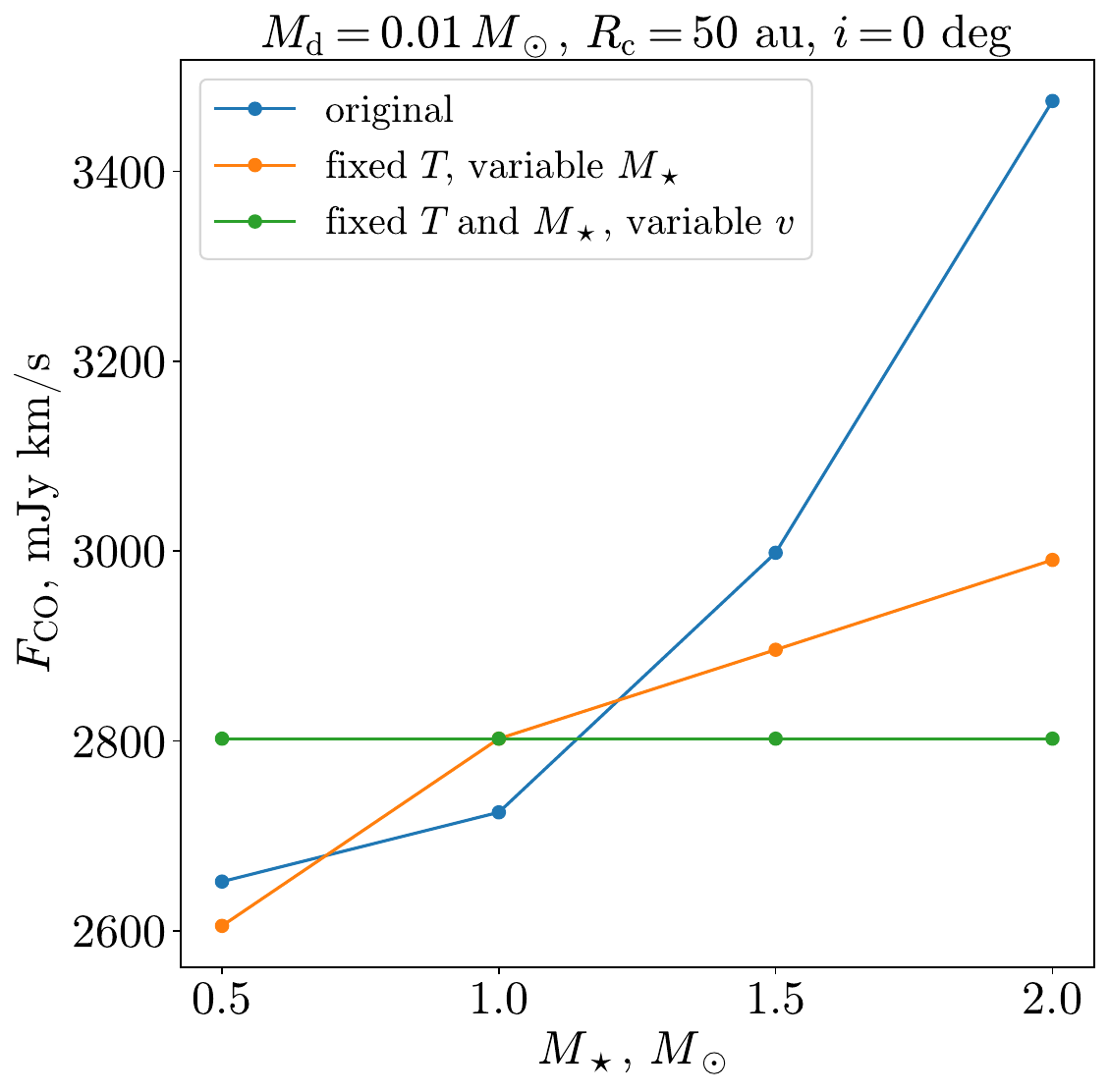}
    \includegraphics[width=0.33\textwidth]{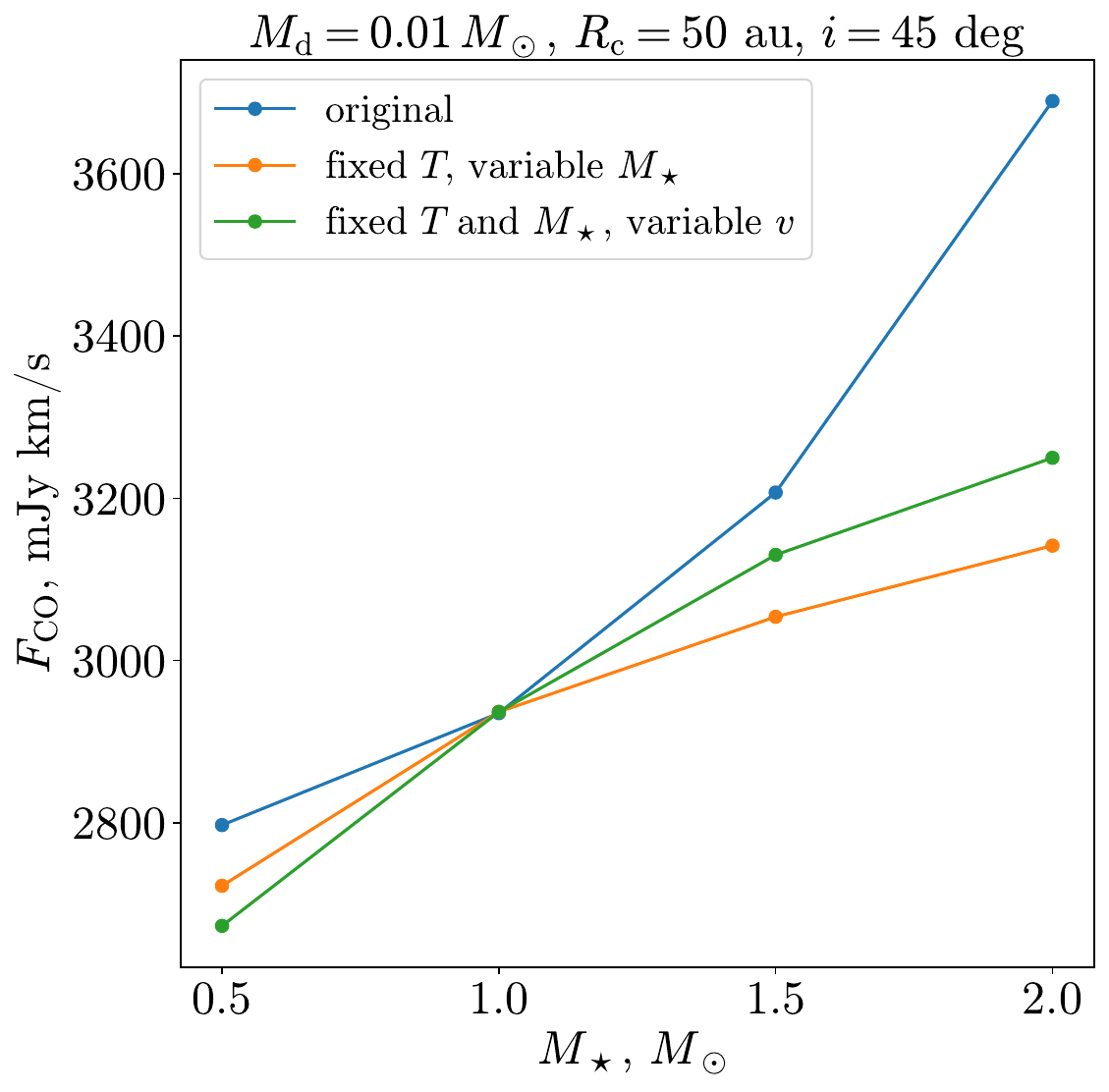}
    \includegraphics[width=0.33\textwidth]{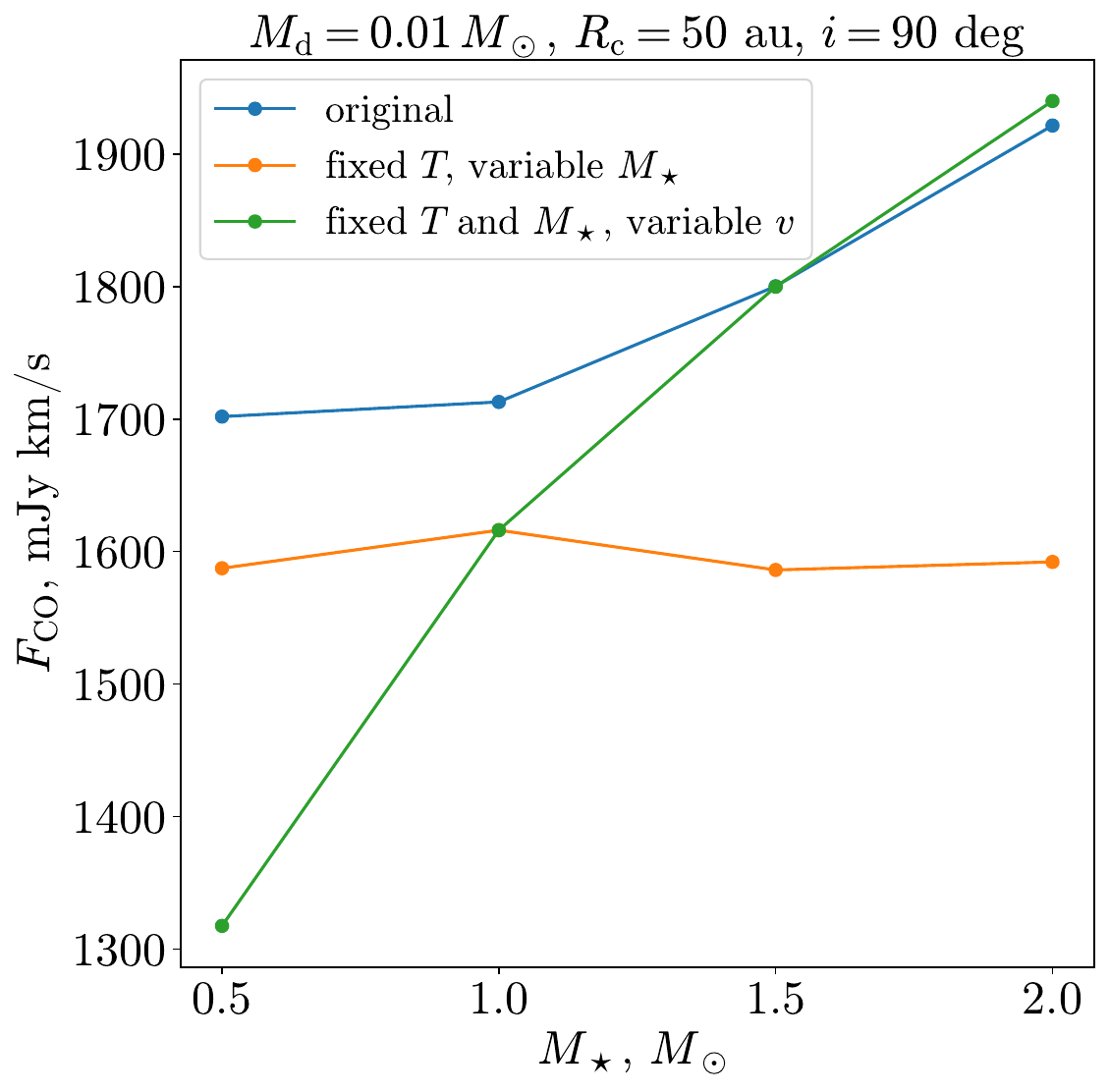}
    \caption{{Dependence of the CO flux on the stellar mass for three inclinations. Left panel: $i=0$\textdegree. Middle panel: $i=45$\textdegree. Right panel: $i=90$\textdegree. Three types of models are considered: original, with {grain surface chemistry}, from Table~\ref{tab:modeldata}; models where temperature and luminosity of the star were fixed to $T_\star = 4000$\,K and $L_\star = 1\,L_\odot$; models where only velocity field was adjusted to stellar mass.}}
    \label{fig:startest}
\end{figure*}

{The effect of stellar mass consists of three contributions: disk thermal structure, vertical distribution of gas, and velocity field. To disentangle these three effects, we consider three sets of models. The first set includes twelve models with $M_{\rm d} = 0.01\,M_\odot$, $R_{\rm c} = 50$\,au, four values of $M_\star$ = 0.5, 1, 1.5, 2$\,M_\odot$ and three inclinations $i$ = 0\textdegree, 45\textdegree, 90\textdegree. These models will experience all the three effects of different stellar mass. For the second set, we run the same twelve models, but with fixed $T_\star=4000$\,K and $L_\star=1\,L_\odot$. This eliminates the effect on the thermal structure, but the vertical density distribution and velocity field in the models are still affected by the stellar mass. Finally, in the third set of models, we simulate the model with $M_\star=1\,M_\odot$, but assume the disk to have velocity distributions corresponding to the four different values of $M_\star$ (at three different inclinations). This set of models will only include the effect of the kinematic structure.}

{Effects of photodissociation are fully included in the ``original'' set of models, both due to the change in stellar temperature (minor) and combined destruction of CO in the molecular layer on dust grain surface and by photodissociation. Effects of photodissociation can also change in the second set of models, as the radiation field distribution will shift according to the density change as the disk scale height decreases, possibly destroying more CO in the upper disk layers.}

{Fig.~\ref{fig:startest} shows the dependences of CO fluxes on stellar mass in these sets of models. First we consider the case of $i=0$\textdegree (left panel). As the disk is face-on and we do not see disk rotation in the line profile, the flux is constant for the third set of models. When we add influence of mass on the vertical density structure, flux starts to increase with star mass ($F(2\,M_\odot)/F(0.5\,M_\odot) = 1.15$) due to a temperature increase along $\tau=1$ surface as the density structure affects the radiation field. Lastly, adding the thermal effect, flux increases more ($F(2\,M_\odot)/F(0.5\,M_\odot) = 1.32$) as the disk surface heats up.}

{For $i=45$\textdegree (middle panel), we start to see the effect on flux from the Doppler shift ($F(2\,M_\odot)/F(0.5\,M_\odot) = 1.2$). Including disk compression now decreases the effect on flux ($F(2\,M_\odot)/F(0.5\,M_\odot) = 1.15$) as a part of the emission surface is the disk outer rim which shrinks with higher stellar mass. Adding the thermal effect raises the flux increase rate further ($F(2\,M_\odot)/F(0.5\,M_\odot) = 1.32$).}

{Finally, in $i=90$\textdegree (right panel) case, Doppler shift has its highest effect ($F(2\,M_\odot)/F(0.5\,M_\odot) = 1.5$). Disk compression, however, completely negates it as now we see only the disk outer rim. In the end we are left with the flux increase from pure thermal effect ($F(2\,M_\odot)/F(0.5\,M_\odot) = 1.13$).}

{Concluding, all three effects give comparable flux change with disk compression acting against the flux increase from rest at higher inclinations and adding to it at lower.}

\section{Parametric vertical structure and no grain surface {reactions}}
\label{app:test_models}

{Fig.~\ref{fig:tau1_test} presents CO gas and ice distributions for four models from Fig.~\ref{fig:lumlumWB14} right panel. The model in the left upper panel is our reference model with {grain surface chemistry} and hydrostatic equilibrium. If we now set the vertical structure parametrically (left lower panel), we see that the $\tau=1$ surfaces shift upwards where temperature is slightly higher which contributes to the flux increase. For C$^{18}$O we also see a slight shift of the $\tau=1$ surface from the star. This can be explained by more matter being above the depletion area as scale height is higher in this case.}

{When we turn off grain surface {reactions} (right upper panel), the region of CO depletion between $T_{\rm freeze}$ and $T_{\rm chem}$ (see Section~\ref{sec:CO_depletion}) becomes rich in CO gas (from 10 to 50\,au in the presented model). This evidently expands the emitting area for $^{13}$CO and C$^{18}$O, thus increasing the flux. Combining both effects (right lower panel) further increases the fluxes.}
\begin{figure*}
    \centering
    \includegraphics[width=\textwidth]{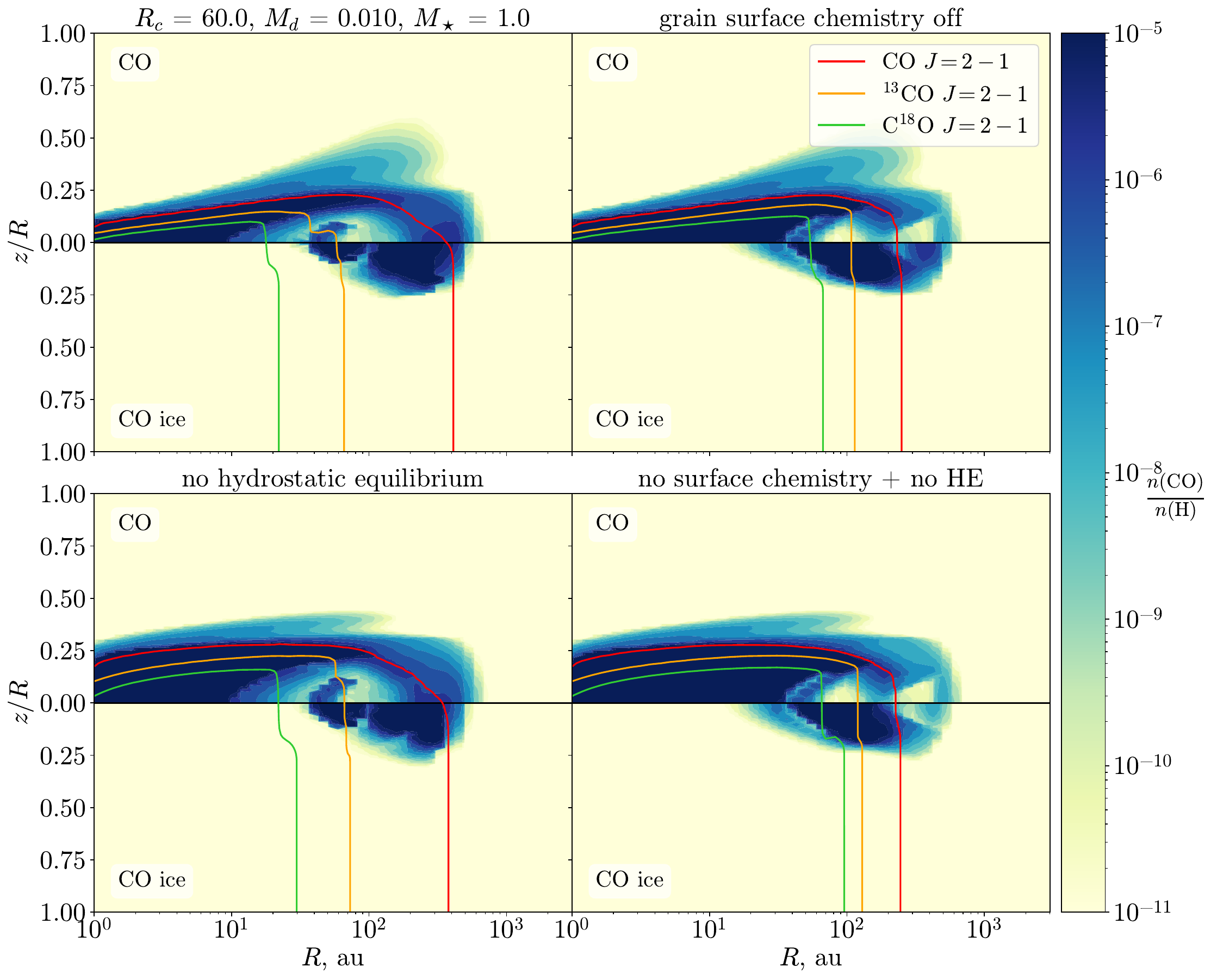}
    \caption{{CO gas and ice abundance distributions in the disk for four models with different assumptions but same parameters. Contours represent $\tau=1$ surfaces of used lines. Left upper panel: a reference model with {grain surface chemistry} and $M_{\rm d} = 0.01\,M_\odot$, $R_{\rm c} = 60$\,au, $M_\star=1.0\,M_\odot$. Right upper panel: no grain surface {reactions}. Left lower panel: parametric vertical structure. Right lower panel: parametric vertical structure and no surface chemistry.}}
    \label{fig:tau1_test}
\end{figure*}
\end{appendix}

\end{document}